\documentclass[twocolumn,letterpaper,aps,prd,superscriptaddress,showpacs,nofootinbib,floatfix]{revtex4}

\usepackage{graphicx}
\usepackage{amssymb}

\usepackage{dcolumn}	

\usepackage{xspace}	

\def\mean#1{\ensuremath{\left<#1\right>}}

\newcommand{\gevc}{\mbox{GeV/$c~$}}

\newcommand{\gevcsq}{\mbox{GeV/$c^{2}~$}}

\newcommand{\pt}{\mbox{$p_T$}~}

\newcommand{\jpsi}{\mbox{$J/\psi~$}}
\newcommand{\jpsib}{\mbox{$J/\psi$}}
\newcommand{\psip}{\mbox{$\psi^{\prime}~$}}
\newcommand{\psipb}{\mbox{$\psi^{\prime}$}}
\newcommand{\ee}{\mbox{$e^{+}e^{-}$~}}

\newcommand{\full}{\mbox{$\sqrt{s}=$ 200 GeV~}}
\newcommand{\fullb}{\mbox{$\sqrt{s}=$ 200 GeV}}
\newcommand{\pp}{\mbox{$p$+$p$~}}

\newcommand{\cc}{\mbox{$c\bar{c}~$}}
\newcommand{\ccb}{\mbox{$c\bar{c}$}}
\newcommand{\bb}{\mbox{$b\bar{b}~$}}
\newcommand{\bbb}{\mbox{$b\bar{b}$}}

\newcommand{\chic}{\mbox{$\chi_c$~}}
\newcommand{\chicb}{\mbox{$\chi_c$}}

\def\func#1{\left(#1\right)}

\begin{document}


\title{Ground and excited charmonium state production 
       in $p$+$p$ collisions at $\sqrt{s}=$ 200 GeV}

\newcommand{\abilene}{Abilene Christian University, Abilene, Texas 79699, USA}
\newcommand{\acadsin}{Institute of Physics, Academia Sinica, Taipei 11529, Taiwan}
\newcommand{\banaras}{Department of Physics, Banaras Hindu University, Varanasi 221005, India}
\newcommand{\barc}{Bhabha Atomic Research Centre, Bombay 400 085, India}
\newcommand{\bnlcoll}{Collider-Accelerator Department, Brookhaven National Laboratory, Upton, New York 11973-5000, USA}
\newcommand{\bnlphys}{Physics Department, Brookhaven National Laboratory, Upton, New York 11973-5000, USA}
\newcommand{\caucr}{University of California - Riverside, Riverside, California 92521, USA}
\newcommand{\charlesczech}{Charles University, Ovocn\'{y} trh 5, Praha 1, 116 36, Prague, Czech Republic}
\newcommand{\chonbuk}{Chonbuk National University, Jeonju, 561-756, Korea}
\newcommand{\ciae}{China Institute of Atomic Energy (CIAE), Beijing, People's Republic of China}
\newcommand{\cns}{Center for Nuclear Study, Graduate School of Science, University of Tokyo, 7-3-1 Hongo, Bunkyo, Tokyo 113-0033, Japan}
\newcommand{\colorado}{University of Colorado, Boulder, Colorado 80309, USA}
\newcommand{\columbia}{Columbia University, New York, New York 10027 and Nevis Laboratories, Irvington, New York 10533, USA}
\newcommand{\czechtech}{Czech Technical University, Zikova 4, 166 36 Prague 6, Czech Republic}
\newcommand{\dapnia}{Dapnia, CEA Saclay, F-91191, Gif-sur-Yvette, France}
\newcommand{\debrecen}{Debrecen University, H-4010 Debrecen, Egyetem t{\'e}r 1, Hungary}
\newcommand{\elte}{ELTE, E{\"o}tv{\"o}s Lor{\'a}nd University, H - 1117 Budapest, P{\'a}zm{\'a}ny P. s. 1/A, Hungary}
\newcommand{\ewha}{Ewha Womans University, Seoul 120-750, Korea}
\newcommand{\fit}{Florida Institute of Technology, Melbourne, Florida 32901, USA}
\newcommand{\fsu}{Florida State University, Tallahassee, Florida 32306, USA}
\newcommand{\gsu}{Georgia State University, Atlanta, Georgia 30303, USA}
\newcommand{\hiroshima}{Hiroshima University, Kagamiyama, Higashi-Hiroshima 739-8526, Japan}
\newcommand{\ihepprot}{IHEP Protvino, State Research Center of Russian Federation, Institute for High Energy Physics, Protvino, 142281, Russia}
\newcommand{\illuiuc}{University of Illinois at Urbana-Champaign, Urbana, Illinois 61801, USA}
\newcommand{\inrras}{Institute for Nuclear Research of the Russian Academy of Sciences, prospekt 60-letiya Oktyabrya 7a, Moscow 117312, Russia}
\newcommand{\instpasczech}{Institute of Physics, Academy of Sciences of the Czech Republic, Na Slovance 2, 182 21 Prague 8, Czech Republic}
\newcommand{\isu}{Iowa State University, Ames, Iowa 50011, USA}
\newcommand{\jinrdubna}{Joint Institute for Nuclear Research, 141980 Dubna, Moscow Region, Russia}
\newcommand{\jyvaskyla}{Helsinki Institute of Physics and University of Jyv{\"a}skyl{\"a}, P.O.Box 35, FI-40014 Jyv{\"a}skyl{\"a}, Finland}
\newcommand{\kek}{KEK, High Energy Accelerator Research Organization, Tsukuba, Ibaraki 305-0801, Japan}
\newcommand{\kfki}{KFKI Research Institute for Particle and Nuclear Physics of the Hungarian Academy of Sciences (MTA KFKI RMKI), H-1525 Budapest 114, POBox 49, Budapest, Hungary}
\newcommand{\korea}{Korea University, Seoul, 136-701, Korea}
\newcommand{\kurchatov}{Russian Research Center ``Kurchatov Institute", Moscow, 123098 Russia}
\newcommand{\kyoto}{Kyoto University, Kyoto 606-8502, Japan}
\newcommand{\labllr}{Laboratoire Leprince-Ringuet, Ecole Polytechnique, CNRS-IN2P3, Route de Saclay, F-91128, Palaiseau, France}
\newcommand{\lawllnl}{Lawrence Livermore National Laboratory, Livermore, California 94550, USA}
\newcommand{\losalamos}{Los Alamos National Laboratory, Los Alamos, New Mexico 87545, USA}
\newcommand{\lpc}{LPC, Universit{\'e} Blaise Pascal, CNRS-IN2P3, Clermont-Fd, 63177 Aubiere Cedex, France}
\newcommand{\lund}{Department of Physics, Lund University, Box 118, SE-221 00 Lund, Sweden}
\newcommand{\maryland}{University of Maryland, College Park, Maryland 20742, USA}
\newcommand{\mass}{Department of Physics, University of Massachusetts, Amherst, Massachusetts 01003-9337, USA }
\newcommand{\muenster}{Institut fur Kernphysik, University of Muenster, D-48149 Muenster, Germany}
\newcommand{\muhlenberg}{Muhlenberg College, Allentown, Pennsylvania 18104-5586, USA}
\newcommand{\myongji}{Myongji University, Yongin, Kyonggido 449-728, Korea}
\newcommand{\nagasaki}{Nagasaki Institute of Applied Science, Nagasaki-shi, Nagasaki 851-0193, Japan}
\newcommand{\newmex}{University of New Mexico, Albuquerque, New Mexico 87131, USA }
\newcommand{\nmsu}{New Mexico State University, Las Cruces, New Mexico 88003, USA}
\newcommand{\ornl}{Oak Ridge National Laboratory, Oak Ridge, Tennessee 37831, USA}
\newcommand{\orsay}{IPN-Orsay, Universite Paris Sud, CNRS-IN2P3, BP1, F-91406, Orsay, France}
\newcommand{\peking}{Peking University, Beijing, People's Republic of China}
\newcommand{\pnpi}{PNPI, Petersburg Nuclear Physics Institute, Gatchina, Leningrad region, 188300, Russia}
\newcommand{\riken}{RIKEN Nishina Center for Accelerator-Based Science, Wako, Saitama 351-0198, Japan}
\newcommand{\rikjrbrc}{RIKEN BNL Research Center, Brookhaven National Laboratory, Upton, New York 11973-5000, USA}
\newcommand{\rikkyo}{Physics Department, Rikkyo University, 3-34-1 Nishi-Ikebukuro, Toshima, Tokyo 171-8501, Japan}
\newcommand{\saispbstu}{Saint Petersburg State Polytechnic University, St. Petersburg, 195251 Russia}
\newcommand{\saopaulo}{Universidade de S{\~a}o Paulo, Instituto de F\'{\i}sica, Caixa Postal 66318, S{\~a}o Paulo CEP05315-970, Brazil}
\newcommand{\seoulnat}{Seoul National University, Seoul, Korea}
\newcommand{\stonybrkc}{Chemistry Department, Stony Brook University, SUNY, Stony Brook, New York 11794-3400, USA}
\newcommand{\stonycrkp}{Department of Physics and Astronomy, Stony Brook University, SUNY, Stony Brook, New York 11794-3400, USA}
\newcommand{\subatech}{SUBATECH (Ecole des Mines de Nantes, CNRS-IN2P3, Universit{\'e} de Nantes) BP 20722 - 44307, Nantes, France}
\newcommand{\tenn}{University of Tennessee, Knoxville, Tennessee 37996, USA}
\newcommand{\titech}{Department of Physics, Tokyo Institute of Technology, Oh-okayama, Meguro, Tokyo 152-8551, Japan}
\newcommand{\tsukuba}{Institute of Physics, University of Tsukuba, Tsukuba, Ibaraki 305, Japan}
\newcommand{\vandy}{Vanderbilt University, Nashville, Tennessee 37235, USA}
\newcommand{\waseda}{Waseda University, Advanced Research Institute for Science and Engineering, 17 Kikui-cho, Shinjuku-ku, Tokyo 162-0044, Japan}
\newcommand{\weizmann}{Weizmann Institute, Rehovot 76100, Israel}
\newcommand{\yonsei}{Yonsei University, IPAP, Seoul 120-749, Korea}
\affiliation{\abilene}
\affiliation{\acadsin}
\affiliation{\banaras}
\affiliation{\barc}
\affiliation{\bnlcoll}
\affiliation{\bnlphys}
\affiliation{\caucr}
\affiliation{\charlesczech}
\affiliation{\chonbuk}
\affiliation{\ciae}
\affiliation{\cns}
\affiliation{\colorado}
\affiliation{\columbia}
\affiliation{\czechtech}
\affiliation{\dapnia}
\affiliation{\debrecen}
\affiliation{\elte}
\affiliation{\ewha}
\affiliation{\fit}
\affiliation{\fsu}
\affiliation{\gsu}
\affiliation{\hiroshima}
\affiliation{\ihepprot}
\affiliation{\illuiuc}
\affiliation{\inrras}
\affiliation{\instpasczech}
\affiliation{\isu}
\affiliation{\jinrdubna}
\affiliation{\jyvaskyla}
\affiliation{\kek}
\affiliation{\kfki}
\affiliation{\korea}
\affiliation{\kurchatov}
\affiliation{\kyoto}
\affiliation{\labllr}
\affiliation{\lawllnl}
\affiliation{\losalamos}
\affiliation{\lpc}
\affiliation{\lund}
\affiliation{\maryland}
\affiliation{\mass}
\affiliation{\muenster}
\affiliation{\muhlenberg}
\affiliation{\myongji}
\affiliation{\nagasaki}
\affiliation{\newmex}
\affiliation{\nmsu}
\affiliation{\ornl}
\affiliation{\orsay}
\affiliation{\peking}
\affiliation{\pnpi}
\affiliation{\riken}
\affiliation{\rikjrbrc}
\affiliation{\rikkyo}
\affiliation{\saispbstu}
\affiliation{\saopaulo}
\affiliation{\seoulnat}
\affiliation{\stonybrkc}
\affiliation{\stonycrkp}
\affiliation{\subatech}
\affiliation{\tenn}
\affiliation{\titech}
\affiliation{\tsukuba}
\affiliation{\vandy}
\affiliation{\waseda}
\affiliation{\weizmann}
\affiliation{\yonsei}
\author{A.~Adare} \affiliation{\colorado}
\author{S.~Afanasiev} \affiliation{\jinrdubna}
\author{C.~Aidala} \affiliation{\mass}
\author{N.N.~Ajitanand} \affiliation{\stonybrkc}
\author{Y.~Akiba} \affiliation{\riken} \affiliation{\rikjrbrc}
\author{H.~Al-Bataineh} \affiliation{\nmsu}
\author{J.~Alexander} \affiliation{\stonybrkc}
\author{A.~Angerami} \affiliation{\columbia}
\author{K.~Aoki} \affiliation{\kyoto} \affiliation{\riken}
\author{N.~Apadula} \affiliation{\stonycrkp}
% ...
\collaboration{PHENIX Collaboration} \noaffiliation


\begin{abstract}

We report on charmonium measurements [$J/\psi$~(1S), 
$\psi^{\prime}$~(2S), and $\chi_c$~(1P)] in $p$+$p$ collisions at 
$\sqrt{s}=$ 200 GeV.  We find that the fraction of $J/\psi$ coming 
from the feed-down decay of $\psi^{\prime}$ and $\chi_c$ in the 
midrapidity region ($|\eta|<0.35$) is 9.6~$\pm$~2.4\% and 
32~$\pm$~9\%, respectively.  New, higher statistics $p_T$ and 
rapidity dependencies of the $J/\psi$ yield via dielectron decay in 
the same midrapidity range and at forward rapidity ($1.2<|\eta|<2.4$) 
via dimuon decay are also reported.  These results are compared with 
measurements from other experiments and discussed in the context of 
current charmonium production models.

\end{abstract}

\date{\today}


\maketitle


\section{Introduction}
\label{section:introduction}

\setcounter{section}{1}\setcounter{equation}{0}

Since its discovery, charmonium (bound \cc states) has been 
proposed as a powerful tool to investigate many aspects of QCD such 
as the distribution of partons in protons and nuclei at large 
momentum transfer. Charm quarks are predominantly produced in gluon 
interactions at \fullb, therefore they are sensitive to the gluon 
distribution in the nucleon and its modification in the nucleus. In 
addition, the color screening of the \cc state makes charmonium 
dissociation an important signature for the formation of a 
deconfined state of matter created in A+A 
collisions\cite{Matsui:1986dk,PhysRevLett.99.211602}. Such studies 
rely on a accurate understanding of charmonium production in \pp 
collisions which is the goal of the present work.

The cross section of \cc production is known from pQCD calculations 
to about a factor of two compared to PHENIX 
data~\cite{PhysRevLett.95.122001,PhysRevLett.97.252002}. However, 
the hadronization step which forms the bound state is a 
nonperturbative process and is not well understood.  A variety of 
schemes have been proposed, some of the most common being the Color 
Evaporation Model (CEM), the Color Singlet Model (CSM) and 
nonrelativistic quantum chromodynamics (NRQCD), also known as the 
Color Octet Model (COM). We review these models briefly here, and 
compare to them later in the text.

In the CEM \cite{Fritzsch:1977ay,Amundson:1996qr} the bound-state production
mechanism is insensitive to the \cc quantum numbers. The \cc pair is
produced as long as the center of mass energy of the pair, $\sqrt{\hat{s}}$, is
greater than the mass of two charm quarks, but less than the mass of
two open charm mesons. Charmonium states are then color neutralized during
the hadronization process by soft gluon emission. The yield of different charmonium states 
is a fixed fraction $\mathcal{F}$ of the integrated pQCD \cc cross
section. $\mathcal{F}$ is determined from experiments and is
universal. Hence in this model, the ratio between the yield of
different charmonium states is momentum and energy independent.

In the CSM \cite{Baier:1981uk} the production amplitude of on-shell
\cc pairs is projected onto $^{2S+1}L_J$ angular momentum
states, and hence accounts for the \jpsi and \psip as $^3S_1$ and the
\chic states as $^3P_{0,1,2}$. The model assumes that these charmonium
states are formed in their final color singlet quantum number
configuration.  The production density matrix is coupled to the wave function at the origin which is determined from
potential models. The
only empirical parameters entering in the entire calculation are the
leptonic decay width and the
charmonium mass used in the potential model.

NRQCD allows for the production of both color singlet and color octet
\cc states. Color octet states emit one or more gluons during
hadronization in order to neutralize their color. The production
amplitude is expanded in powers of both the strong coupling,
$\alpha_S$, and the velocity, $\nu$, of the heavy quarks relative to
the \cc pair. The expansion in $\nu$ assumes that the heavy quark is
nonrelativistic.\footnote{Potential model calculations indicate the
  velocity of charm(bottom) is $\sim$ 0.23(0.1)$c$.} As in the CSM,
the production amplitudes are projected onto $^{2S+1}L_J$ states. Since
the potential model can only be applied to the color singlet state, a
long range nonperturbative matrix element for each quarkonium state
is fitted from experiments.  The earliest such
matrix parametrization \cite{Cho:1995ce} was tuned with \jpsi and
\psip cross sections observed in CDF \mbox{($\sqrt{s}$ = 1.8 TeV)}
\cite{Abe:1997jz} which indicated that while P-wave charmonium (\chic)
has no important color octet state contributions, S-wave charmonium
(direct \jpsi and \psip) production is largely through color
octet channels. Therefore, this model is sometimes simply referred to as the
Color Octet Model (COM).

Each model has its strengths and weaknesses.
The CEM is able to reasonably describe quarkonia yields observed
in many experiments, but has no predictive power for \cc polarization.
Cross sections calculated using CSM grossly underestimate the yields
observed at PHENIX \cite{Adare:2006kf} and at CDF
\cite{Abe:1997jz}. Recent next-to-leading order (NLO)
\cite{Campbell:2007ws,Artoisenet:2007xi,Gong:2008sn} and
next-to-next-leading order (NNLO)
\cite{Artoisenet:2008fc,Lansberg:2008gk,Lansberg:2009db,Lansberg:2010vq}
calculations for the color singlet states resulted in significant
modifications of the predicted charmonium yields and polarization, but
not sufficient to agree with the experimental results. NRQCD tuned
with \jpsi and \psip \pt spectra from CDF was able to qualitatively
describe the first PHENIX \jpsi cross section and polarization results
\cite{PhysRevD.82.012001} albeit with large experimental
uncertainties, but failed to describe the \jpsi and \psip polarization
observed in CDF \cite{Abulencia:2007us} (see \cite{Lansberg:2008gk}
for a recent review). Recent NRQCD calculations
\cite{Butenschoen:2010rq} include color singlet and color octet NLO
short range terms along with a long-range matrix parametrization
from experimental hadroproduction \cite{PhysRevD.71.032001} and
photoproduction \cite{Adloff:2002ex,Aaron:2010gz} of \jpsi
mesons. However, the NLO terms for the color octet \cite{Gong2009197}
have only small corrections compared to the leading-order (LO) terms
and the calculations still disagree with the \jpsi polarization measured by CDF.

One of the complications in the \jpsi total cross section and
polarization calculations (observables where experimental tests are
readily available) is the contribution from the decays of excited
charmonium states, primarily \psipb, $\chi_{c1}$ and $\chi_{c2}$. In
addition, the \jpsi suppression observed in heavy
ion collisions cannot be completely understood without a knowledge of
the feed-down fraction of excited charmonium state decays to the \jpsib.  This
is particularly true under the assumption that the suppression is due
to the disassociation of charmonium in the a high temperature
quark-gluon plasma, since lattice calculations
\cite{PhysRevLett.99.211602} indicate that the melting points of the
\chic and \psip states are lower than that of the \jpsi.  In this work the
feed-down fractions to the \jpsi from excited charmonium states is
measured since they can be determined more precisely than
production cross sections as many of the systematic uncertainties
cancel when making cross section ratios.

The PHENIX Experiment at RHIC can measure quarkonia dilepton decays
over a broad \pt and rapidity range and can detect photons from \chic
radiative decays using electromagnetic calorimeters at
midrapidity. This article reports on the feed-down fraction of
the \jpsi yield which comes from \psip and \chic decays at
midrapidity in \pp collisions at \fullb. For these we used the \psip
to \jpsi yield ratio in the dielectron channel and the full
reconstruction of the $\pp \rightarrow \chic \rightarrow \jpsi +
\gamma \rightarrow e^+e^- + \gamma$ decay. A new \jpsi differential
cross section measurement at midrapidity and forward rapidity using the
increased luminosity obtained in the 2006 and 2008 runs, is also
presented. These provide more accurate measurements than previously
published in \cite{Adare:2006kf}, particularly for the \jpsi
differential cross section at high \pt.  The results obtained in these
analyses also provide a baseline for the study of \jpsi suppression in
$d$+Au \cite{Adare:2010fn} and Au+Au \cite{Adare:2011yf,Adare:2006ns} collisions at PHENIX.

Systematic uncertainties throughout this article are classified
according to whether or not there are point-to-point correlations
between the uncertainties. Type A systematic uncertainties are
point-to-point uncorrelated, similar to statistical uncertainties,
since the points fluctuate randomly with respect to each other. Type B
systematic uncertainties are point-to-point correlated. The points
fluctuate coherently with respect to each other. That is, it accounts
for the uncertainty in the nth-order derivative of the measured
spectrum, in most cases the slope.  Global, or type C, systematics are
those where all points fluctuate in the same direction and by the same
fractional amount.

The remainder of the article is arranged as follows. An introduction
to the PHENIX detector, a description of the data sample, and a description
of the
lepton identification method is described in Sec. \ref{sec:apparatus}.
The analysis is described in three sections: midrapidity
\jpsi and \psip dielectron cross section measurement in the PHENIX
central arms (Sec. \ref{sec:jpsi_psip_analysis}); direct \chic
feed-down measurement in the central arms
(Sec. \ref{sec:chic_analysis}); and forward rapidity \jpsi
cross section measurement in the muon arms
(Sec. \ref{sec:dimuon_analysis}). The results are compared to
measurements from other experiments and to current theoretical
calculations in Sec. \ref{sec:results_discussion}.

\section{Experimental Apparatus and the Data Set}
\label{sec:apparatus}

The PHENIX detector \cite{Adcox:2003zm} is composed of four arms. Two
central arms measure electrons, photons and hadrons over $|\eta|<0.35$
with each azimuthally covering $\Delta\phi=\pi/2$. Two forward muon arms
measure muons over the range $-2.2<\eta<-1.2$ arm and $1.2<\eta<2.4$
with full azimuthal coverage.

Charged particle tracks in the central arms are formed using the Drift
Chamber (DCH), the Pad Chamber (PC) and the collision point. Electron
candidates required at least one fired phototube within an annulus
\mbox{3.4 $< R_{ring} [cm] <$ 8.4} centered on the projected track
position on the Ring Imaging \v{C}erenkov detector (RICH).\footnote{Corresponding to $\Delta \phi = 8$mrad and $\Delta
  Z=3$cm} In addition, the electron candidate is required to be
associated with an energy cluster in the Electromagnetic Calorimeter
(EMCal) that falls within 4$\sigma_{position}$ of the projected track
position, and within 4$\sigma_{energy}$ of the expected
energy/momentum ratio where the $\sigma$'s characterize the position
and energy resolution of the EMCal. The relatively loose association
requirement still provides excellent hadron rejection due to the very
small particle multiplicity in \pp collisions. Based on the \pt range
of decay electrons from \jpsi observed in real data and simulations, a
minimum $p_T$ of 500 MeV/$c$ was also required for each electron
candidate.

Each of the forward muon arms\cite{Akikawa:2003zs} comprises a
hadron absorber, three stations of cathode strip chambers for particle
tracking (MuTr), and a Muon Identifier detector (MuID). The hadron
absorber is composed of a 20 cm thick copper nosecone and 60 cm of
iron which is part of the magnet. The MuTr is installed in an
eight-sided conical magnet. The MuID is composed of five steel hadron
absorbers interleaved with six panels of vertical and
horizontal Iarocci tubes.  A single muon needs a longitudinal momentum
of 2 \gevc to reach the most downstream MuID plane.
Tracks reconstructed in the MuTr are identified as muons if they match
a ``road'' formed by hits in the MuID, within 2.5$\sigma$ of angular
resolution. At least one tube in the last MuID plane 
should have fired. Additional cuts include a $\chi^2<23$ for the
reconstructed track, a $\chi^2<9$ for the track projection to the
collision vertex and a polar angle cut of $14^\circ<\theta_{\mu}<33^\circ$ for
the north arm and $147^\circ<\theta_{\mu}<166^\circ$ for the south arm to avoid
acceptance inconsistencies between the detector simulation and real
data near the edges of the muon arms.

Beam interactions were selected with a minimum-bias (MB) trigger that
requires at least one hit per beam crossing in each of the two
beam-beam counters (BBC) placed at $3.0<|\eta|<3.9$.  Studies using
Vernier scans (also called van der Meer scans)
\cite{PhysRevLett.91.241803} conclude that this MB trigger accepts a
cross section of $\sigma_{\rm BBC}$ = 23.0 $\pm$ 2.2 mb. This cross
section represents 55 $\pm$5\% of the \mbox{$\sigma_{pp}^{inel.}$ = 42 $\pm$
3 mb} \pp inelastic cross section at \fullb.

Dedicated triggers were used to select events with at least one
electron or two muon candidates. An EMCal RICH
Trigger (ERT) required a minimum energy in any 2$\times$2
group of EMCal towers, corresponding to $\Delta\eta \times
  \Delta\phi = 0.02 \times 0.02$ rad., plus associated hits in the
RICH. The minimum EMCal energy requirement was 400 MeV for the
first half of the Run and 600 MeV for the second half. The data used in this
analysis were taken with the ERT in coincidence with the
MB trigger. Events were also triggered when there were two muon
candidates in one of the MuID arms. The trigger
logic for a muon candidate required a ``road'' of fired Iarocci tubes in
at least four planes, including in the most downstream plane
relative to the collision point. The event sample used in the
dimuon analysis required a MuID trigger in coincidence with the MB
trigger.

There are events which produce a \jpsi but do not fire the MB
trigger. The fraction, $\varepsilon_{inel.}$,of such events is estimated by measuring the
number of high $p_T$ $\pi^0\rightarrow \gamma\gamma$ decays which
satisfy the minimum energy condition of the ERT and which do not
satisfy the MB trigger.  It was found that $\varepsilon_{inel.} = (79
\pm 2)\%$.  The correction due to this factor is included in all cross
section calculations for measurements requiring the MB trigger. No
dependence of $\varepsilon_{inel.}$ on the $p_T$ of the measured
$\pi^0$ decays was found over the range 0-10 \gevc\cite{Adare:2010de}.

The collision point along the beam direction is determined with a
resolution of 1.5 cm by using the difference between the fastest time
signals measured in the north and south BBC detectors.  The collision 
point was required to be within $\pm$30 cm of the nominal center of the
detector.  In the dielectron analysis, runs in which electron yields
were more than three standard deviations away from the average in at
least one of the eight EMCal azimuthal sectors, were discarded. 
For the dimuon analysis, runs where the muon arm spectrometers were not
fully operational were rejected. 

The 2006 data sample used in the dielectron analysis corresponded to
$N_{pp} = $ 143 billion minimum bias events, or an integrated luminosity of
$\int \mathcal{L} dt = N_{pp}/\sigma_{\rm BBC} = (6.2 \pm 0.6)$ pb$^{-1}$.
The 2006 and 2008 data samples used for the muon analysis,
corresponded to 215 billion minimum bias events, or a luminosity
of $(9.3 \pm 0.9)$ pb$^{-1}$.

\section{\jpsi and \psip analysis in the midrapidity region}
\label{sec:jpsi_psip_analysis}

The procedure for analyzing the \jpsi and \psip $\rightarrow$
dielectron signal in the central arm detectors is detailed in this
section. The overall procedure to select dielectrons and extract the
charmonium signal and determine combinatorial and correlated backgrounds is
explained in \ref{sec:dielectron_id}. Studies of the central arm
detector response to charmonium dielectron decays is the subject of
the section \ref{sec:dielectron_acceptance}. The final \pt and
rapidity dependence of the cross sections is calculated in Section
\ref{sec:jpsi_psip_cross_section} together with a summary of all
systematic uncertainties mentioned throughout the text. Finally the
\psipb/(\jpsi) dielectron yield ratio is calculated in Section
\ref{sec:psip_jpsi_ratio}.

\subsection{Di-electron decays of \jpsi and \psip mesons in the
  midrapidity region.}
\label{sec:dielectron_id}

\begin{figure}
  \centering
  \includegraphics[width=1.0\linewidth]{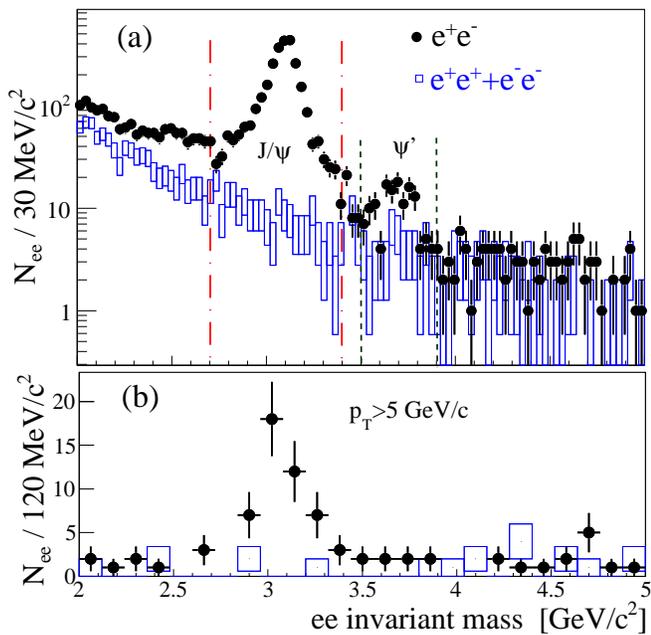}
  \caption{\label{fig:jpsi_peak}(Color online) Invariant mass
    distribution of unlike-sign (closed circles) and like-sign (open boxes)
    dielectrons in the \jpsi and \psip region without (a) and with
    (b) a minimum \pt requirement for the dielectron pair. 
    Dash-dotted (dashed)
    lines represent the mass range used to count \jpsi (\psip) decays.}
\end{figure}

The invariant mass was calculated for all electron pairs in which one
electron of the pair geometrically matched the position of a fired ERT
segment. This requirement was necessary given that we used simulated
\jpsi and \psip decays to estimate the ERT efficiency.  Di-electron
contributions to \jpsi and \psip decays are clearly identified as
peaks in this invariant mass distribution (Fig. \ref{fig:jpsi_peak}).
The primary sources of physically correlated unlike-sign pairs
($e^+e^-$) are quarkonia decays, open \cc and \bb pairs, Drell-Yan,
and jets. Uncorrelated unlike-sign pairs are from combinatorial
background. The primary sources of like-sign pairs ($e^+e^+ + e^-e^-$)
are combinatorial background,
and electrons from particle decays
occurring in the same jet (mostly $\pi^0$ Dalitz decays). The
like-sign pair mass distribution normalized by the geometric mean of
the number of $e^+e^+$ and $e^-e^-$ pairs was statistically subtracted
from the unlike-sign mass distribution.  The primary effect of this
subtraction was to account for combinatorial background, however it
also accounted for much of the jet background.  There are 2882
unlike-sign and 203 like-sign dielectrons in the \jpsi mass range
\mbox{(2.7$<M_{ee} [\gevcsq] <$3.4)}, giving a correlated signal of
2,679 $\pm$ 56 counts and a signal/background of 13.  In the \psip
mass region (3.5$<M_{ee} [\gevcsq] <$3.9) there were 137 unlike-sign
and 51 like-sign electron pairs corresponding to a signal of 86 $\pm$
14 counts and signal/background of 1.7.

The jet contribution in the charmonium mass region is three orders of
magnitude smaller than from the \jpsi and \psip with a steeply falling
mass spectrum\cite{PhysRevC.81.034911} and will be ignored here; in
any case it is largely removed by the like-sign subtraction.  The
Drell-Yan contribution was estimated using next-to-leading-order
calculations \cite{Vogelsang:2007}. Taking into account the detector
acceptance, the fraction of the dielectron signal which comes from
Drell-Yan processes is 0.23 $\pm$ 0.03 \% in the \jpsi mass region and
\mbox{3.37 $\pm$ 0.40 \%} in the \psip region.  The heavy quark
contribution is the major background to the correlated dielectron
spectrum.  In fact, they represent a significant fraction of the
correlated dielectrons in the \psip mass region.  They will be
estimated by two models as described in the next several paragraphs.

In order to understand the dielectron spectrum, a simulation was done
for the three primary contributions to the mass spectrum: the \jpsi
and \psipb, heavy quark pairs, and Drell-Yan.  The first step was to
generate the initial correlated electron pair spectrum.  The \jpsi and
\psip were generated by weighting their distributions in order to
obtain the same \pt spectrum as seen in real data.  The \jpsi
radiative decay ($\jpsi \rightarrow \ee + \gamma$), also called
internal radiation, was introduced using the mass distribution
estimated from QED calculations \cite{Spiridonov:2004mp}. Drell-Yan
pairs were generated according to the mass distribution obtained from
NLO calculations.  In order to make a conservative estimate and
determine whether the result is model independent, the \cc and \bb
mass distributions were obtained using two different methods:

\bf{1. A dielectron generator:} \rm The semi-leptonic heavy flavor
yield measured in \cite{Adare:2010de,Adare:2006hc} was split into the
\cc ($d\sigma_{c\bar{c}}/dp_T$) and \bb ($d\sigma_{b\bar{b}}/dp_T$)
distributions according to the $c/b$ ratio from fixed-order plus
next-to-leading-log (FONLL) calculations \cite{PhysRevLett.95.122001}
which agree with PHENIX measurements of separated \cc and \bb
production\cite{PhysRevLett.103.082002,Aggarwal:2010xp}. These \cc and \bb yields were
used as input for an electron Monte Carlo generator with uniform
rapidity distribution ($|y|<0.5$) and the measured vertex
distribution. An electron and positron from the decay of a heavy quark
pair were generated for each event.  In this method the heavy quarks
are assumed to have no angular correlation.

\bf{2. {\sc pythia}:} \rm Hard scattering collisions were simulated using
the {\sc pythia}\cite{Sjostrand:2006za} generator. Leading order pair
creation sub-processes and next-to-leading-order flavor creation and
gluon splitting sub-processes are all included in the heavy quark
generation \cite{Norrbin:2000zc}. These sub-processes have different
opening angles for the heavy quark pair. The simulation used the
CTEQ6M \cite{CTEQ6} parton distribution functions (PDF), a Gaussian
$k_T=$ distribution of width 1.5 \gevc, a charm quark mass of 1.5 GeV
and bottom quark mass of 4.8 GeV. Variations of the $k_T$ distribution
and masses of the heavy quarks were included in the systematic
uncertainties. The $p_T$ dependence of electrons from \cc and \bb
given by the simulation agrees with the PHENIX measurement of single
electrons from heavy flavor decay\cite{Adare:2010de}.

The generated electron pairs from all sources were then used as input
to a {\sc geant}-3 \cite{GEANT} based detector Monte-Carlo which included
effects such as Bremsstrahlung radiation of electrons when crossing
detector material and air (external radiation).  Simulated events were
then reconstructed and analyzed using the same criteria as were used
for real data and reported in Sec.\ref{sec:apparatus} and
Sec.\ref{sec:dielectron_id}. More details will be given later in
Sec.\ref{sec:dielectron_acceptance}, including methods of estimating
systematic errors.

The resulting simulated distributions were then fit to the 2-dimensional
mass vs \pt distribution of the measured dielectron signal in the
mass range \mbox{$2.0<M_{e^+e^-} [\gevcsq] <8.0$}. The fit parameters
included the normalization of \cc, \bb, \jpsi and \psip contributions,
the fraction of the internal radiation, and a mass resolution correction
for the simulated resonance peaks. The normalization of the Drell Yan
was fixed according to expectations from the NLO calculations.

\begin{figure*}
\begin{minipage}{1.0\linewidth}
  \centering
  \includegraphics[width=0.55\linewidth]{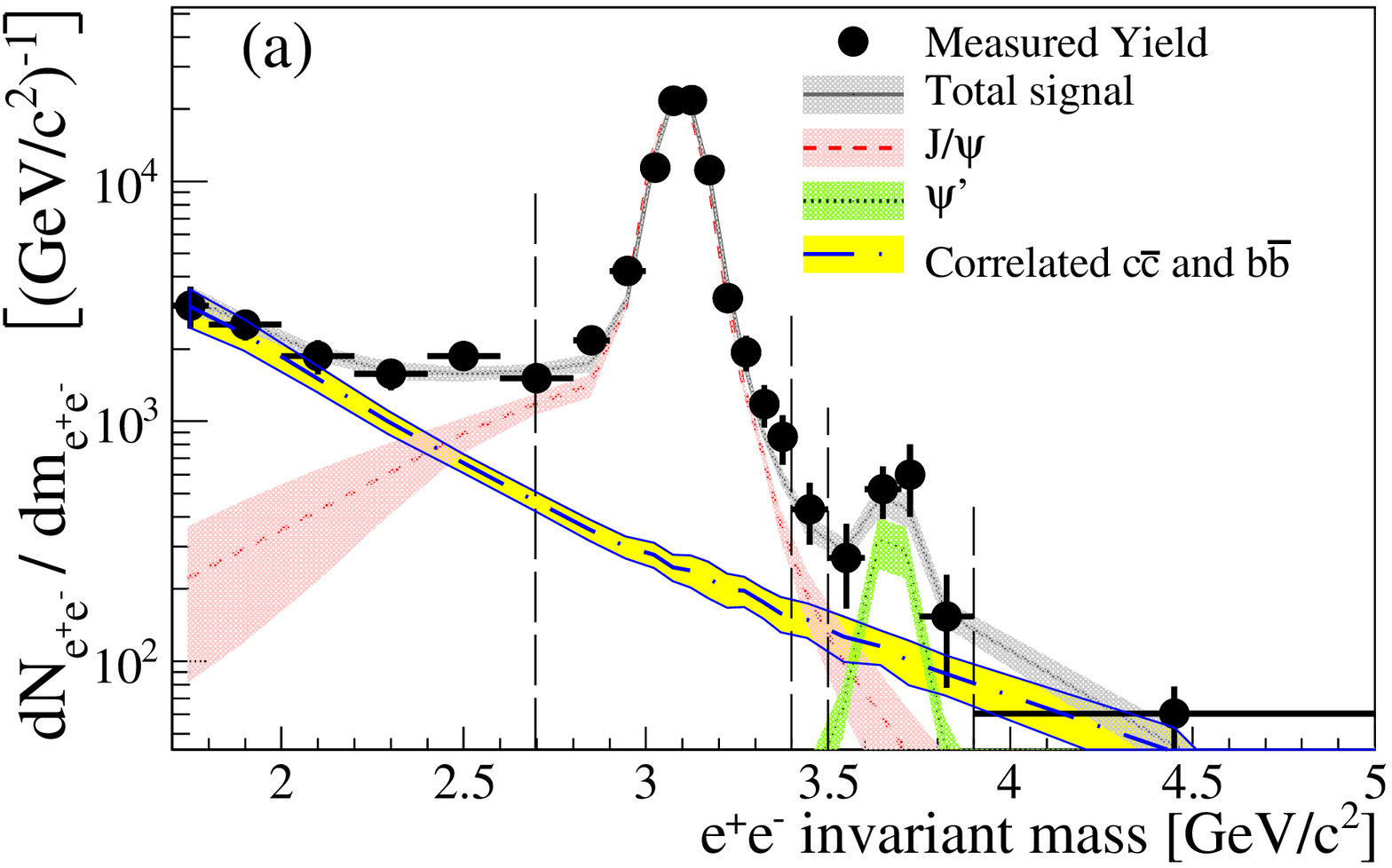}\\
  \includegraphics[width=0.33\linewidth]{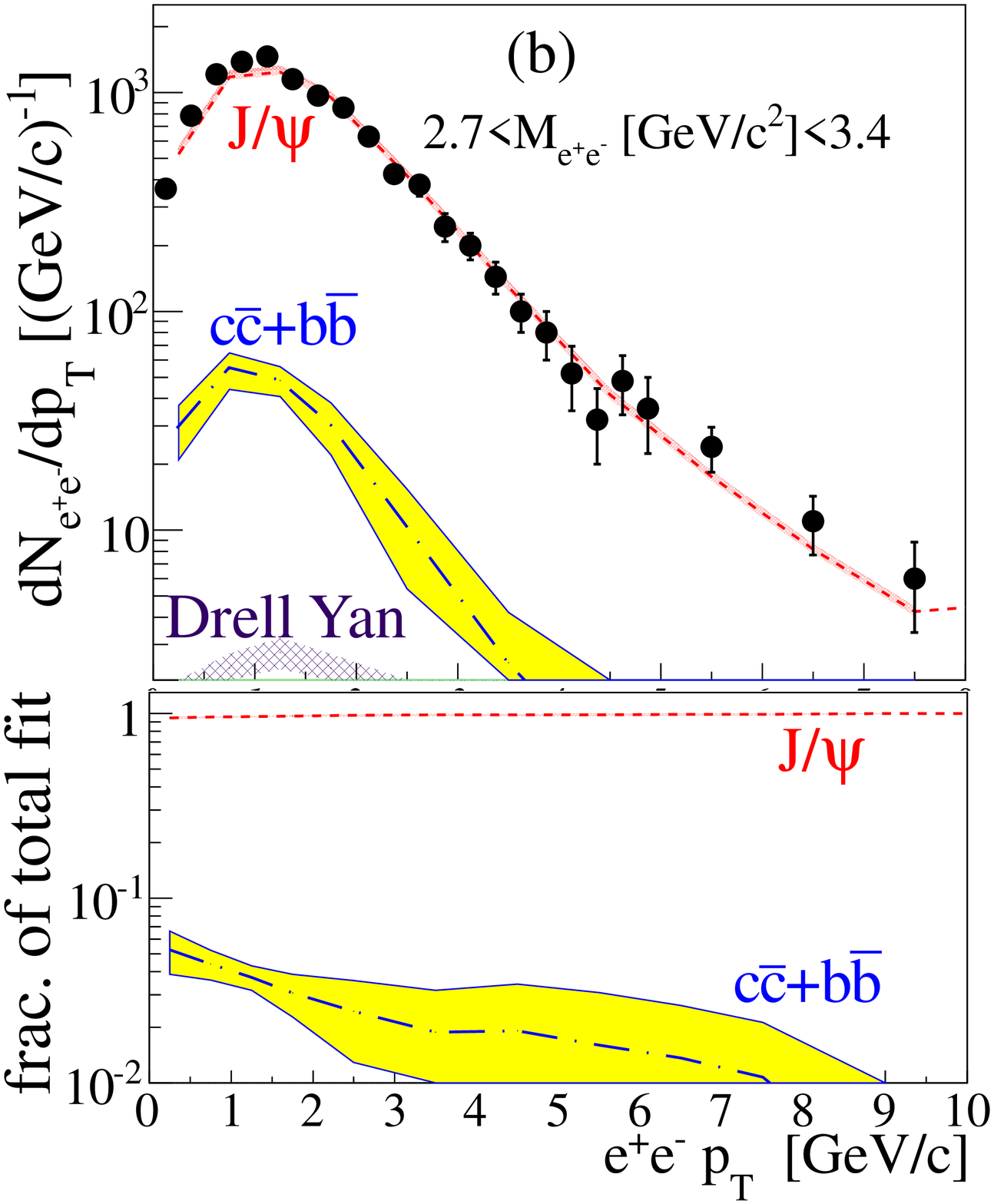}
  \includegraphics[width=0.33\linewidth]{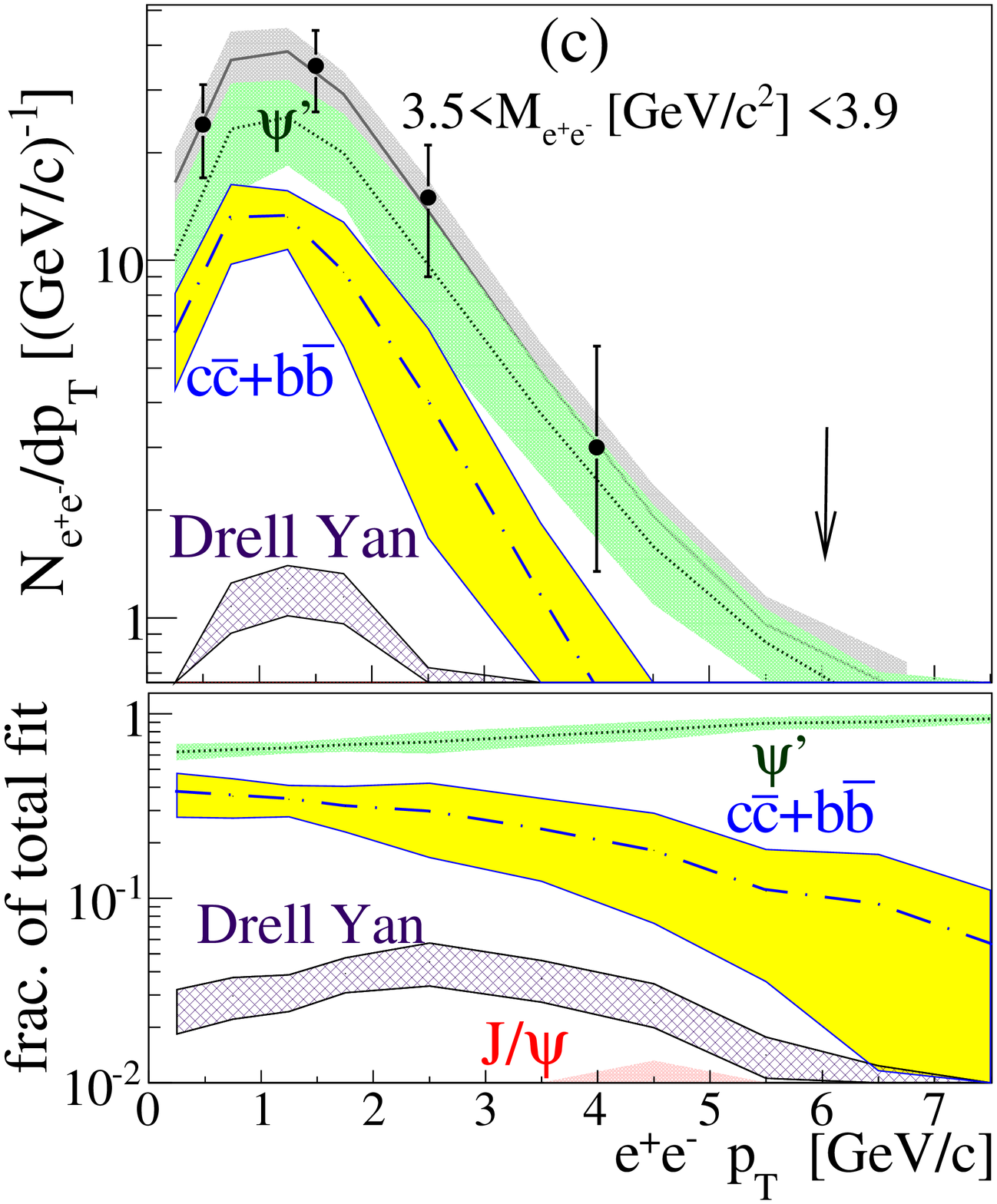}
  \includegraphics[width=0.32\linewidth]{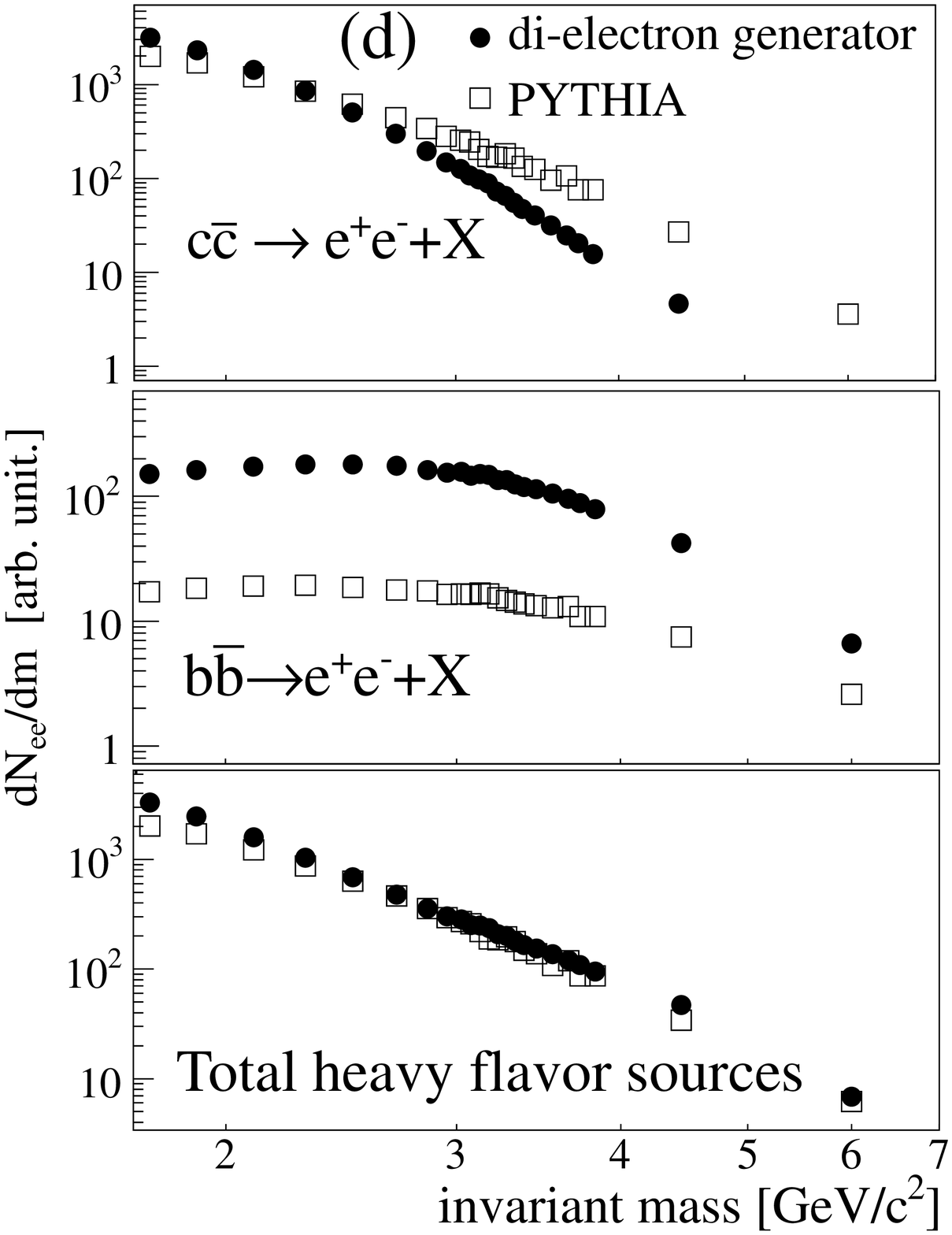}
  \caption{\label{fig:invmass_fit}(Color online) Correlated
    dielectron mass (a) and \pt distributions in the \jpsi (b) and
    \psip (c) mass regions. Signal components were estimated by
    fitting open heavy flavor, Drell Yan (normalization fixed by NLO
    calculations), \jpsi and \psip decays after detector
    simulation. The \cc and \bb components were generated using {\sc pythia}
    \cite{Sjostrand:2006za} and a heavy flavor based dielectron
    generator described in the text. Bands correspond to the type A
    fitting uncertainties and the type B systematic uncertainty
    obtained when using the two different open heavy quark
    generators. Panel (d) shows the result after the fit for \ccb, \bb
    and total open heavy flavor components from each generator.}
\end{minipage}
\end{figure*}

Fig. \ref{fig:invmass_fit} shows the results of the fit for the
dielectron mass (a) and \pt (b,c).  The heavy flavor contribution to
the continuum obtained from the fit using the dielectron generator
and {\sc pythia} is shown in Fig. \ref{fig:invmass_fit}-d. When using the
{\sc pythia} simulation, the presence of back-to-back correlated \cc and \bb
pairs produced more high mass pairs per \cc which then forced
a smaller contribution from \bbb. 
As can be seen from the figure, the fits performed using
the two generators give very different
normalizations for the open charm and the open bottom
contributions. However, the two methods give very similar
contributions for the sum which is well constrained by data. 
Thus the lack of the knowledge of the
angular correlation in heavy flavor production does not affect the
estimate of the total continuum contribution from open heavy
flavor in the \jpsi and \psip mass regions.  The measurement of the
\cc and \bb cross sections is not in the scope of this paper; a more
detailed study can be found in
\cite{:2008asa,Adare:2009ic,Aggarwal:2010xp}.  Type A fit parameter
uncertainties and the type B uncertainty obtained from the difference
in results obtained using the two generators for the total heavy
flavor contribution, are summed in quadrature and shown as bands in
Fig. \ref{fig:invmass_fit}. Values for the fraction of the charmonium signal
$\left(f_{\psi} \right)$ shown in Figs. \ref{fig:invmass_fit}-b and
\ref{fig:invmass_fit}-c are used later in the yield calculation.

The fitted external and internal radiation contributions indicate that
the fraction of radiative decays of the \jpsi, where the undetected
photon has energy larger than 100 MeV, is \mbox{(9 $\pm$ 5)\%}. This
is consistent with QED calculations which indicate that 10.4\% of the
dielectron decays from the \jpsi come from such radiative decays and
a measurement of fully reconstructed $J/\psi \rightarrow e^+e^-\gamma$
performed by E760 \cite{PhysRevD.54.7067} which gives 14.7 $\pm$ 2.2
\%. The \jpsi mass peak around 3.096 \gevcsq has a Gaussian width from
the fit of \mbox{53 $\pm$ 4 MeV} after including a mass resolution in the MC
of ($\delta M/M$) of (1.71 $\pm$ 0.13)\%. Because of the radiative
tails, the mass range \mbox{(2.7$<M_{ee} [\gevcsq] <$3.4)} contains
\mbox{$\varepsilon_{mass}^{J/\psi}=93.8 \pm 0.9$\%} of the \jpsi
decays and the mass region \mbox{(3.5$<M_{ee} [\gevcsq]<$3.9)}
contains \mbox{$\varepsilon_{mass}^{\psi'}=86 \pm 2$\%} of the \psip
decays, corrections included in the yield calculations.
The foreground yield as well as the statistical uncertainties used in
the cross section calculations were obtained assuming that both
foreground and background distributions are independent and follow
Poisson statistics. The total foreground was then multiplied by the
factors obtained from fits in the previous section to obtain the \jpsi
and \psip yields.  In each bin of \pt (or {\it y}) the foreground
signal $(\mu_f)$ was obtained from the unlike-sign counts $(fg)$ in
the distribution and the background $(\mu_b)$ was obtained from the
like-sign counts $(bg)$ (Fig. \ref{fig:jpsi_peak} top).  The joint
probability distribution for the net number of counts $s=\mu_f-\mu_b$
is

\begin{equation}
  P(s,\mu_b) = \frac{\mu_b^{bg}}{bg!}
  \frac{\mu_b^{fg}}{fg!}e^{-2\mu_b}
  \left(1 + \frac{s}{\mu_b} \right)^{fg} e^{-s}. \end{equation}

\noindent  We expand the term $\left(1 + \frac{s}{\mu_b}
\right)^{fg}$:

\begin{eqnarray}
  \left(1 + \frac{s}{\mu_b} \right)^{fg} &=& \sum_{k=0}^{fg}
  \frac{fg!}{(fg-k)!k!}\left(\frac{s}{\mu_b}\right)^k \\\nonumber
  P(s,\mu_b) &=& \sum_{k=0}^{n}
  \frac{\mu_b^{bg+fg-k}e^{-2\mu_b}}{bg!(fg-k)!}\frac{s^ke^{-s}}{k!}. \\\nonumber \end{eqnarray}

\noindent  Assuming no negative signal, the expression is summed over $\mu_b$ from 0 to $\infty$ using the normalization of the Gamma distribution

\begin{equation}
  \int_{0}^{\infty} dx x^{p-1}e^{-bx} = \frac{p-1}{b^p} \end{equation} and $b=2$, $p-1=bg+fg-k$. We obtain finally,

\begin{eqnarray}
  \label{eq:tannembaum_prob}
  P(s) &=&
  \sum_{k=0}^{fg}\frac{(bg+fg-k)!}{bg!(fg-k)!)}\frac{1}{2}\left(\frac{1}{2}
  \right)^{bg+fg-k}\frac{s^ke^{-s}}{k!}.
\end{eqnarray}

%
%
%

The number of charmonium decays for each bin, and
the corresponding statistical uncertainty, were obtained using
(\ref{eq:tannembaum_prob}) given the fraction $\left(f_{\psi}
\right)$ of charmonium in the sample found previously.

\begin{eqnarray}
  \label{eq:tannembaum}
  N_{\psi} &=& \mean{s} \times f_{\psi}.
\end{eqnarray}

\subsection{Di-electron acceptance and efficiency studies}
\label{sec:dielectron_acceptance}

\begin{figure}
  \centering
  \includegraphics[width=1.0\linewidth]{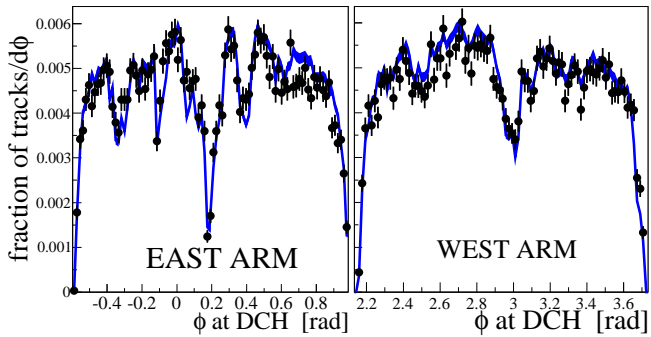}
  \caption{\label{fig:acceptance_phi}(Color online) Simulated (solid
    line) and real data (points) single electron  
    distributions in the $\phi$ coordinate of the drift
    chamber. Error bars correspond to statistical uncertainties.} 
\end{figure}

The detector response to \jpsi and \psip dielectron decays was
studied using the {\sc geant}-3 based Monte Carlo
simulation. Malfunctioning detector channels were removed from the
detector simulation and from the real data analysis. The geometric
acceptance of the detector Monte Carlo was compared to that for real
data using simulated $\pi^0$ decays. The majority of the electrons
found in real data come from $\pi^0$ Dalitz decays and photons which
convert to electrons in the detector structure. The simulated
electrons from $\pi^0$ decays were weighted in order to match the
collision vertex and \pt distributions observed in the data. Fig.
\ref{fig:acceptance_phi} shows the simulated and real electron track
distribution as a function of the azimuthal angle, $\phi$, measured at the
DCH radius. The ratio between real and simulated track distributions
($f_{acc}(\phi_{DCH},z_{DCH})$) is used later to estimate the systematic
uncertainty of the \jpsi acceptance.

\begin{figure} 
\centering
  \includegraphics[width=1.0\linewidth]{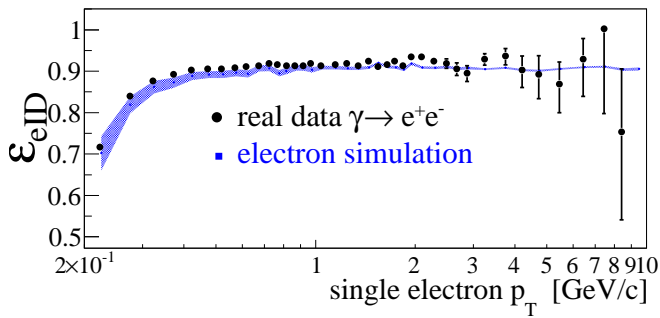}
  \caption{\label{fig:eid_eff} Single electron identification
    efficiency estimated using photon conversions from
    real data (points) and the electron simulation (shaded area).}
\end{figure}

The electron identification efficiency was estimated using
$\gamma \rightarrow \ee$ conversions coming primarily from the beam
pipe.
These
dielectrons, which do not originate from the event vertex, have a nonzero
 invariant mass and can be identified since their invariant mass exhibits
a peak in the region below 30 MeV$/c^2$. Assuming
all tracks in the peak above the combinatorial background are
electrons, the electron identification efficiency was obtained from
the fraction of dielectron conversions which survive the
identification criteria applied to both electron and positron compared
to the number of dielectron conversions obtained after requiring
identification for only one electron or positron. The same procedure
was repeated in the simulation.  Fig. \ref{fig:eid_eff} shows the
electron identification efficiency as a function of the \pt of the
electron in question.  The difference in efficiency between simulation
and data for electrons with $p_T>0.5$ \gevc was no larger than 0.8\%,
which translated to an overall type B uncertainty in the dielectron
yield of 1.1\% due to our understanding of the electron identification
efficiency.

\begin{figure} 
\centering
  \includegraphics[width=1.0\linewidth]{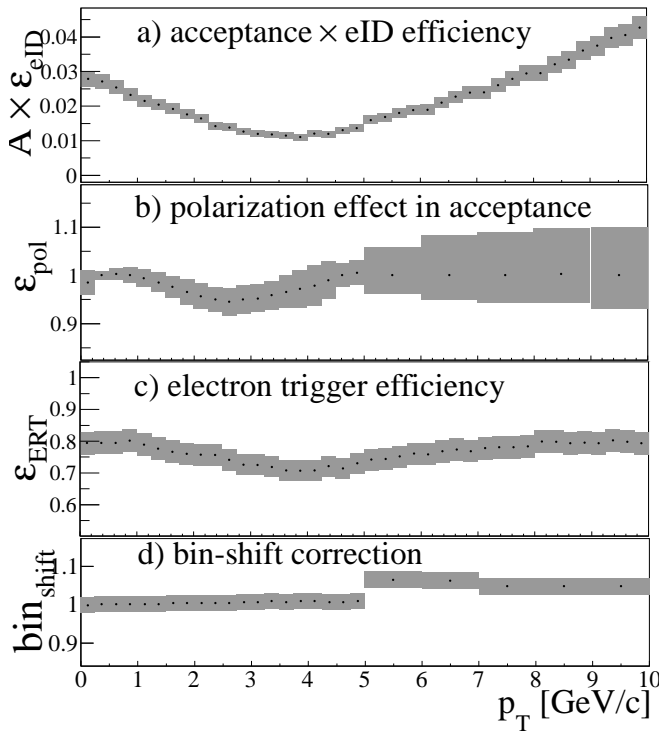}
  \caption{\label{fig:jpsi_efficiency} Transverse momentum dependence
    of the detector performance and correction factors for dielectron
    decays of \jpsi mesons in $|y|<0.5$. Shaded bands are the
    uncertainties of the estimates as described in the text.}
\end{figure}

Simulated \jpsi dielectron decays were generated with uniform \pt and
rapidity ($|y|<0.5$) and the measured vertex distribution. The
fraction of the generated \jpsi decays that were fully reconstructed
corresponds to the acceptance $\times$ electron identification
efficiency of the detector ($A\times\varepsilon_{eID}$) for \jpsi
dielectron decays with rapidity $|y|<0.5$ (Fig.
\ref{fig:jpsi_efficiency}-a).  When each simulated electron decay was
weighted according to $f_{acc}(\phi_{DCH},z_{DCH})$ given previously,
the number of reconstructed \jpsi decays was modified by 7.5\%. This
is essentially the variation in our acceptance calculation, when calculated
using a data driven method as compared to simulation.  We considered
this deviation as a type B systematic uncertainty. The
$A\times\varepsilon_{eID}$ for simulated \psip dielectron decays in
the same rapidity range was larger than that from the \jpsi by between
5-20\% because of its larger mass.  The maximum difference occurs at
$p_T\sim$2.5 \gevc.

The detector acceptance for charmonium also depends on the orientation of
its electron decay with respect to the momentum direction of the 
parent particle, an outcome of charmonium polarization. The
correction factor from polarization ($\varepsilon_{pol}$) was
evaluated using a measurement of \jpsi polarization \cite{PhysRevD.82.012001} 
in \pp collisions interpolated to the relevant transverse momentum.
The uncertainty in $\varepsilon_{pol}$ due to the uncertainty in the
polarization was assigned as a type B systematic uncertainty. In
the \pt region where there is no polarization  measurement
($p_T>5$ \gevc for \jpsi and all \pt for \psip) the one standard deviation
uncertainty was calculated assuming the \jpsi polarization in this
region can be anything between -1 and 1.  Fig.
\ref{fig:jpsi_efficiency}-b shows the \pt dependence of 
$\varepsilon_{pol}$.

The trigger (ERT) performance was studied using single electrons. We
used a MB data sample to measure the \pt dependent fraction of
electron candidates that fired the ERT in each of the EMCal
sectors. These fractions were then used in simulation to estimate the
\jpsi efficiency of the ERT trigger ($\varepsilon_{ERT}$).  This
process was repeated for each change in the ERT operational
conditions, such as a change in the energy threshold, or a significant
modification in the number of EMCal or RICH sectors included in the
ERT trigger.  Fig. \ref{fig:jpsi_efficiency}-c shows the \pt
dependence of $\varepsilon_{ERT}$, weighted by the luminosity
accumulated in each ERT period. When the single electron ERT
efficiency of each EMCal sector was varied within its statistical
uncertainty, a one standard deviation change of 4.5\% in
$\varepsilon_{ERT}$ was observed. This deviation is shown in Fig.
\ref{fig:jpsi_efficiency}-c as the shaded band and is assigned as a
type B systematic uncertainty for the \jpsi and \psip yields. No
significant change in $\varepsilon_{ERT}$ was observed if one used the \psip
in the simulations.

A final correction ($bin_{shift}$) was made for the dominance of the
yield in the lower end of each \pt bin
(Fig. \ref{fig:jpsi_efficiency}-d).
In addition, a correction of up to 2\% was made to account
for bin-by-bin smearing effects due to finite momentum
resolution($bin_{smear}$).

\subsection{Cross section results}
\label{sec:jpsi_psip_cross_section}

The \jpsi and \psip dilepton differential cross section for each \pt
bin is calculated by

\begin{eqnarray}
\label{eq:cross_section}
\frac{B_{ll}^{\psi}}{2\pi p_T} \frac{d^2\sigma_{\psi}}{dydp_T} &=&
B_{ll}^{\psi} \frac{1}{2\pi
  p_T}\frac{d^2N/dydp_T}{\varepsilon_{inel.}\int
  \mathcal{L}dt}\\\nonumber 
\frac{d^2N}{dp_Tdy} &=& \frac{N_{\psi}}{\Delta y \Delta p_T A\varepsilon},
\end{eqnarray}

\noindent where $B_{ll}^{\psi}$ is the branching ratio of the
charmonium states into dileptons and \mbox{$\varepsilon =
\varepsilon_{eID}\varepsilon_{ERT}\varepsilon_{pol}\varepsilon_{mass}bin_{shift}bin_{smear}$}. 

All systematic uncertainties described in the previous sections are
listed and classified in Table \ref{tab:sys_error}. The quadratic sum
of the correlated systematic uncertainties (type B) is between 10\%
and 13\% of the measured \jpsi yield and between 12\% and 22\% of the
measured \psip yield, depending on \pt.

\begin{table}[!ht]
  \caption{\label{tab:sys_error} List of the systematic uncertainties
    relative to the \jpsi and \psip dielectron yields. Ranges indicate \pt dependence.}
  \begin{ruledtabular}\begin{tabular}{lcc}
    description & contribution & type\\\hline
    fraction of \jpsi in the mass cut & 0.4\% & A\\
    fraction of \psip in the mass cut & 3-13\% & A\\
    acceptance & 7.5\% & B \\
    eID efficiency & 1.1\% & B\\
    mass cut efficiency for \jpsi & 1.0\% & B\\
    mass cut efficiency for \psip & 2.0\% & B\\
    heavy flavor MC used in fit for \jpsi & 0.5-1.1\% & B\\
    heavy flavor MC used in fit for \psip & 4.8-10\% & B\\
    up-in-down bin correction & 3\% & B\\
    momentum smear effect & 1.5\% & B\\
    \pt, $y$ and vertex input in $\psi$ MC & 2.0\% & B\\
    \jpsi polarization bias in acceptance & 0-10\% & B\\
    \psip polarization bias in acceptance & 4-17\% & B\\
    ERT efficiency & 4.5\% & B \\
    luminosity & 10\% & C\\
\end{tabular}\end{ruledtabular}
\end{table}

\begin{figure}
  \centering
  \includegraphics[width=1.0\linewidth]{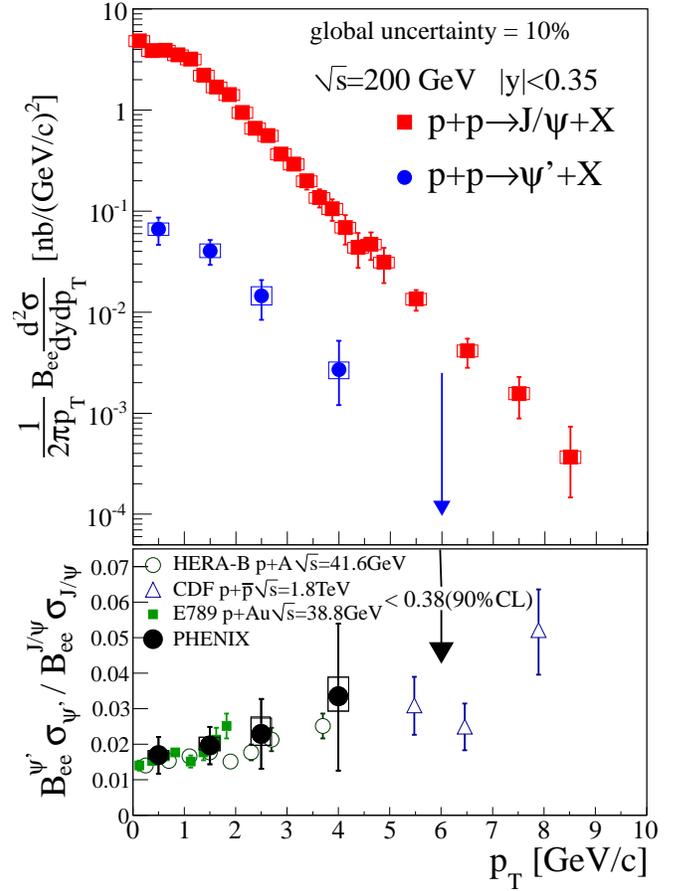}
  \caption{\label{fig:jpsi_psip_yield} Transverse momentum dependence
    of \jpsi and \psip yields in $|y|<0.35$ (top). \psip/(\jpsi) ratio together with
    ratios obtained in other experiments (bottom). Error bars reflect statistical and
type A uncertainties while boxes reflect the quadratic sum of type B uncertainties).}
\end{figure}

The \pt dependencies of the measured \jpsi and \psip yields are shown in Fig.
\ref{fig:jpsi_psip_yield}(top) and Tables \ref{tab:jpsi_yield},
\ref{tab:psip_yield}. The bars correspond to the quadratic sum of all
type A and statistical uncertainties. Boxes represent the quadratic
sum of the type B uncertainties. There is a global uncertainty (type C) of
10\%.

The \pt integrated \jpsi cross section 
was calculated for three rapidity ranges using 

\begin{eqnarray}
  B_{ll}\frac{d\sigma}{dy} = \sum_{p_T}
  B_{ll}\frac{d^2\sigma}{dp_Tdy} \Delta p_T  
\end{eqnarray}

\noindent where $B_{ll} d^2\sigma/dp_Tdy$ is obtained from
(\ref{eq:cross_section}) using $A\varepsilon_{eID}$,
$\varepsilon_{ERT}$ and $\varepsilon_{pol}$ recalculated for each
of the three rapidity bins. The results 
are listed in Table \ref{tab:jpsi_rap} and shown in Fig.
\ref{fig:jpsi_rap}.

\subsection{\psipb/(\jpsi) yield ratio and fraction of  \jpsi
  yield coming from  \psip decays.}
\label{sec:psip_jpsi_ratio}

The decay of \psip to \jpsi cannot be measured in
the current detector configuration. However, we can calculate the
fraction of \jpsi coming from \psip decays $\func{F_{\psi'}^{J/\psi}}$
using the ratio between the \psip and \jpsi 
cross sections and the \psip branching ratio to \jpsi
($B_{J/\psi}^{\psi'} = (58.7 \pm 0.8)$\% \cite{Amsler:2008zzb}).

\begin{eqnarray}
\label{eq:psip_feed-down}
F_{\psi'}^{J/\psi} &=& \frac{B_{J/\psi}^{\psi'} \sigma_{\psi'}}{\sigma_{J/\psi}}.
\end{eqnarray}

We start from the ratio between the \psip and the \jpsi dielectron
counts $R_{J/\psi}^{\psi'}$.
Its joint probability distribution is calculated from the expected
Poisson probability distributions (Eq. \ref{eq:tannembaum_prob}) 
$P_{\psi'}(s_{\psi'})$ and $P_{J/\psi}(s_{J/\psi})$ for the dielectron
counts in the \psip and \jpsi mass ranges respectively, and the
corresponding values $f_{\psi'}$
and $f_{J/\psi'}$ which account for the 
fraction of \psip and \jpsi contributions in the chosen dielectron
mass ranges:

\begin{eqnarray}
  P\left(R_{J/\psi}^{\psi'}\right)  = \frac{P_{\psi'}(s_{\psi'})f_{\psi'}}{P_{J/\psi}(s_{J/\psi})f_{J/\psi}}.
\end{eqnarray}

The \psipb/(\jpsi) dielectron cross section ratio is thus determined
as follows where the different correction factors for \psip and \jpsi 
must be taken into account.

\begin{eqnarray}
\frac{B_{e^+e^-}^{\psi'}\sigma_{\psi'}}{B_{e^+e^-}^{J/\psi}\sigma_{J/\psi}}
= \mean{R_{J/\psi}^{\psi'}} \frac{(A\varepsilon_{eID})^{J/\psi}\varepsilon_{ERT}^{J/\psi}\varepsilon_{mass}^{J/\psi}\varepsilon_{pol}^{J/\psi}}
  {(A\varepsilon_{eID})^{\psi'}\varepsilon_{ERT}^{\psi'}\varepsilon_{mass}^{\psi'}\varepsilon_{pol}^{\psi'}}.
\end{eqnarray}

Type A uncertainties are propagated for $f_{\psi'}$ and $f_{J/\psi'}$ 
while common relative type B uncertainties that are
correlated for \jpsi and \psip cancel. The remaining
uncertainty in the ratio comes from the quadratic difference between
type B uncertainties which are different for the \jpsi and \psipb. The
\psipb/(\jpsi) dielectron cross section ratio is shown in the bottom
panel of Fig.  \ref{fig:jpsi_psip_yield}. The numbers are listed in
Table \ref{tab:psip_jpsi_ratio}.

Using the branching ratios, $B_{e^+e^-}^{\psi'} = (0.765 \pm 0.017)\%$ 
and $B_{e^+e^-}^{J/\psi} = (5.94 \pm 0.06)\%$
\cite{Amsler:2008zzb} in (\ref{eq:psip_feed-down}) gives
\begin{eqnarray}
  \label{eq:psip_feeddown}
  F_{\psi'}^{J/\psi} = (9.6 \pm 2.4)\%. 
\end{eqnarray}

\section{Radiative decay of \chic}
\label{sec:chic_analysis}

The decay channel $\chic \rightarrow \jpsi + \gamma \rightarrow \ee +
\gamma$ is fully reconstructed in the central arms and is used to
directly measure the feed-down fraction of \chic decays in the
inclusive \jpsi yield ($F_{\chi_c}^{J/\psi}$). This measurement is
particularly challenging since the photon is typically of very low
energy. The data sample used in this
measurement and the $\gamma$ identification procedure is
described in Section \ref{sec:chic_selection}. The detector
performance for the measurement of photon decays of the \chic is
discussed in Section \ref{sec:chic_acceptance}. The composition of all
combinatorial and correlated backgrounds for the \chic signal in the
$\ee\gamma$ mass distribution is detailed in Section
\ref{sec:chic_signal}. Section \ref{sec:chic_result} presents the
final feed-down fraction calculation and a summary of all
uncertainties.

\subsection{Selection of $\chic \rightarrow \jpsi + \gamma$ decays}
\label{sec:chic_selection}

The analysis of the radiative decay of the \chic requires the
identification of photons with energy ($E_{\gamma}$) as low as 300
MeV, the lower limit of the energy we allow in this analysis. Photons
were identified as energy clusters in the EMCal whose profile is
consistent with an electromagnetic shower.  This profile is based on
the response of the EMCal to electron beam tests performed before the
EMCal installation \cite{Aphecetche:2003zr}.  Energy clusters that
were closer than four standard deviations (of the energy cluster
position resolution) to reconstructed charged tracks were rejected, in
order to remove electron and misidentified hadron
contributions. Electrons from photon conversions in detector material
which were not reconstructed by the tracking system were removed by
requiring energy clusters to be further than four standard deviations
from hits in the Pad Chamber (PC) located in front of the EMCal.

The invariant mass of $e^+e^- + \gamma$ is formed using \ee pairs in a
tight \jpsi mass region of $2.9<M_{e^+e^-} [\gevcsq]<3.3$, avoiding
the region where photons produced by Bremsstrahlung radiation can
become an additional background in the 300 MeV energy region.  The
sample contains $N_{J/\psi}=$2456 $\pm$ 51 \ee pairs from \jpsi
decays, after removing combinatorial and correlated background as done
previously. The $e^+e^-\gamma$ mass distribution is plotted in Fig.
\ref{fig:massdist_jpsigamma} (top), where we require E$_{\gamma}>$300
MeV.  The mass of $\ee\gamma$ minus the mass of the measured \ee pair
is plotted in order to cancel the effect of the mass resolution in the
\ee pair. The remaining resolution in the subtracted mass distribution
is from the energy resolution of the measured photon.

\subsection{Detector performance for  \chic radiative decay}
\label{sec:chic_acceptance}

\begin{figure}
  \centering
  \includegraphics[width=1.0\linewidth]{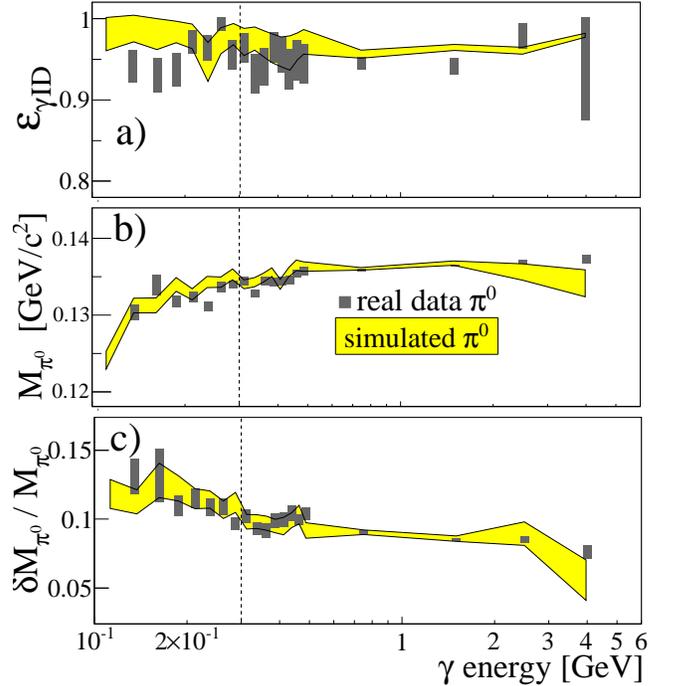}
  \caption{\label{fig:pi0_resolution} (Color online)Study of the
    $\pi^0$ detection performance using both the measured $\gamma$
    energy in real data (boxes) and
    detector MC (shaded band). (a) $\gamma$ identification efficiency,
    (b) $\pi^0$ mass peak position, (c) $\pi^0$ mass
    resolution. The vertical dashed line represents the minimum $\gamma$
    energy required in the \chic analysis. Uncertainties are from the
    $\pi^0$ fit parameters in simulated and real data.}
\end{figure}

The resolution of the mass distribution $M_{e^+e^-\gamma} -
M_{e^+e^-}$ is dominated by the photon energy resolution of the
EMCal. Most photons from \chic decays have energy close to the lower
limit of the EMCal sensitivity.  The behavior of the calorimeter was
studied by using a clean sample of $\pi^0 \rightarrow \gamma\gamma$
decays in real data and in the simulations. Pairs of clusters were
formed where the invariant mass of the pair was required to be
consistent with a $\pi^0$.  Only one of the clusters was required to
pass electromagnetic shower requirements.  The photon identification
efficiency was obtained assuming the other cluster of the pair was
a photon.  This was done on a statistical basis by subtracting a
mixed event background to account for the small contamination from
random clusters under the $\pi^0$ peak.
Fig. \ref{fig:pi0_resolution}-a shows the energy dependence of the
photon identification efficiency ($\varepsilon_{\gamma ID}$) obtained
using real and simulated $\pi^0$s.  The simulation gives an efficiency
2.3\% larger than that found in real data. This difference was
assigned as a type B systematic uncertainty in $\varepsilon_{\gamma
  ID}$.

The central value of the $\pi^0$ mass peak decreases slightly as the
photon energy approaches the lower limit of the calorimeter
sensitivity.  This behavior is caused by zero suppression during
data acquisition and the energy cluster recognition algorithm.
These effects are correctly reproduced in simulation as can be seen in Fig.
\ref{fig:pi0_resolution}-b.  The $\gamma$ energy resolution ($\delta
E_{\gamma}/E_{\gamma}$) was uniformly degraded by 4.7\% in the simulation
in order to match the mass resolution ($\delta M/M$) of the $\pi^0$
peaks observed in real data (Fig. \ref{fig:pi0_resolution}-c).

\begin{figure}[!ht]
  \centering
  \includegraphics[width=1.0\linewidth]{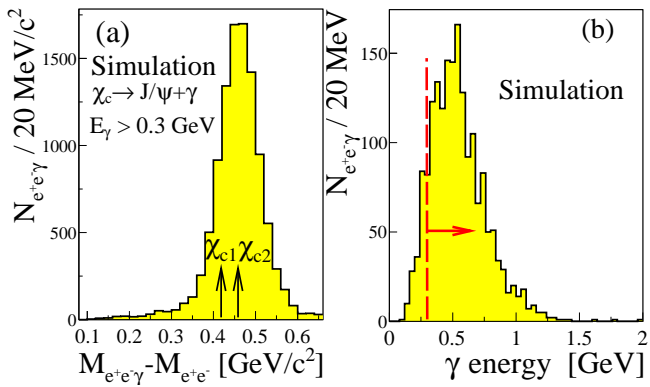}
  \caption{\label{fig:chic_eegamma}(Color online) Mass (a) and $\gamma$
    energy (b) distributions of $\ee\gamma$ decays from $\chi_{c1}$ and
    $\chi_{c2}$ decays obtained from {\sc pythia} + detector
    simulation. The dashed line in (b) represents the $\gamma$ energy cut
    applied in this analysis.}
\end{figure}

\begin{figure}[!ht]
  \centering
  \includegraphics[width=1.0\linewidth]{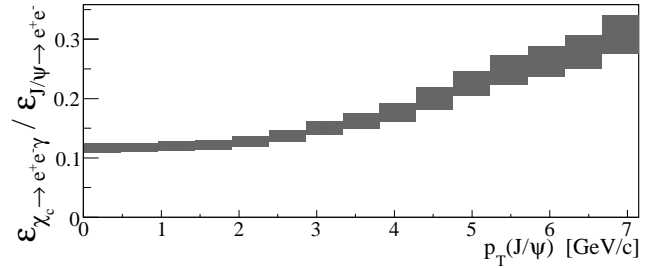}
  \caption{\label{fig:chic_eff}Conditional acceptance and efficiency
    of the \chic decay $\gamma$ as a function of the \jpsi transverse
    momentum. The height of the boxes corresponds to the type B
    systematic uncertainty due to the lack of knowledge of the
    polarization and the photon identification efficiency.}
\end{figure}

$\chi_{c1}$ and $\chi_{c2}$ states were generated using gluon+gluon
scattering in {\sc pythia} with the CTEQ6M PDF, requiring that the \jpsi be
in the rapidity range $|y|<0.5$. The $\chi_{c0}$ is not considered in
the simulation because of its small branching ratio to \jpsi of (1.14
$\pm$ 0.08)\% \cite{Amsler:2008zzb}. Fig. \ref{fig:chic_eegamma}
shows the mass and $\gamma$ energy distribution of $\ee\gamma$ decays
of simulated \chicb. The conditional acceptance of $\gamma$ from \chic
is plotted as a function of the \jpsi momentum in Fig.
\ref{fig:chic_eff}. The detector geometric acceptance of the \chic can
be affected by its polarization and the polarization of the decay
\jpsi. There is no measurement of the \chic polarization. Simulation
studies found the overall acceptance is modified by at most 5.6\% if
the \chic is totally transversely polarized. This possible
modification was included in the acceptance type B systematic
uncertainty.

\subsection{Composition of the $\ee + \gamma$ sample.}
\label{sec:chic_signal}

In addition to the \chic signal, the observed $\ee + \gamma$ sample is
composed of combinatorial background, mostly coming from uncorrelated
$\pi^0$ decays present in events where a \jpsi is detected, and by
photonic sources correlated to the \jpsi which will be discussed
later.

The combinatorial background from random \ee pairs (i.e. the
combinatorial background to the \jpsi in the \chic decay) is well
described by the sum of $e^+e^+\gamma$ and $e^-e^-\gamma$ mass
distributions. This sum was normalized by the geometrical average of
the two components, and subtracted from the $\ee\gamma$ mass spectrum.
The mass distribution of random $(\ee)+\gamma$ combinations
(i.e. essentially random \jpsi+$\gamma$ pairs) was obtained using the
invariant mass distribution of \ee pairs from one event
and photons another.
In order to obtain the combinatorial background as
realistically as possible, events used to form the \ee and $\gamma$
combination were required to have event vertices within 3 cm
(2$\sigma$ of the vertex position resolution) of each other.

The sources of correlated background include internal and external
(Bremsstrahlung) radiative decays of \jpsi, i.e. $\jpsib\rightarrow\ee
\gamma$, $\pi^0$s produced in jets containing \jpsi,
$\psipb\rightarrow\jpsi + $ neutral mesons, $B^0 \rightarrow \jpsi +$
X where X or its decays includes a $\gamma$. Another possibility is
that a \jpsi could be produced together with a high energy
photon\cite{Sridhar:1992ny}. Recent studies also suggest an important
contribution from $gg \rightarrow \jpsi+\gamma+gg$ in NNLO
calculations at $\sqrt{s}=$14 TeV \cite{Lansberg:2009db}. No estimate
was made for \full at the time of this writing. These sources will be
considered in the next few paragraphs.

Photons produced by Bremsstrahlung radiation in the detector structure
are very close to their associated electron and are rejected by the
criteria that removes electrons in the $\gamma$ identification. The
minimum dielectron mass cut of 2.9 \gevcsq also removes radiative
\jpsi decays with $E_{\gamma}>200$ MeV, i.e. those in the energy range
of the photons used in this analysis.

\begin{figure}[!ht]
  \centering
  \includegraphics[width=1.0\linewidth]{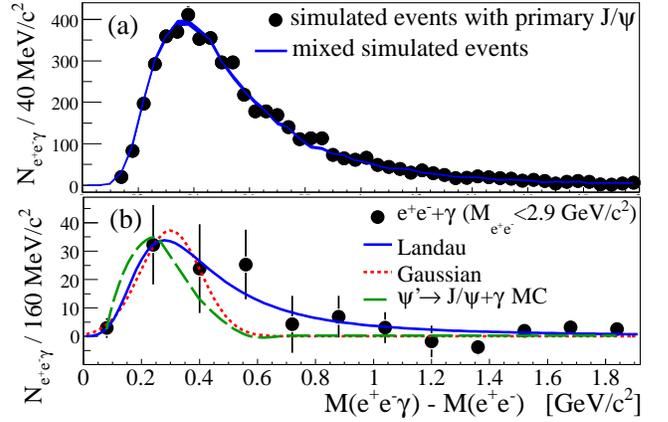}
  \caption{\label{fig:pythia_jpsi_gamma_sim} (Color online) (a) Simulated
    $e^+e^- +\gamma$ invariant mass distribution from {\sc pythia} events
    containing primary \jpsi decays. The line is the combinatorial
    background obtained using simulated mixed events from the same
    sample (top). (b) The $e^+e^-\gamma$ mass distribution in the data, after combinatorial background subtraction,
    where $M_{e^+e^-} [\gevcsq]<$2.9.  The lines are empirical fits as
    explained in the text.  Note that the simulated $\psip \rightarrow
    \jpsi+\gamma$ shape is arbitrarily normalized.}
\end{figure}

Collisions containing primary \jpsi mesons produced by gluon+gluon
scattering (the dominant source) were simulated using {\sc pythia} in order
to understand the electron radiation and jet contributions. Only the
\ee and the radiative $\ee\gamma$ decay channels were allowed. All
final state particles with momentum larger than 100 MeV and
$|\eta|<0.5$ were reconstructed. $\jpsi$ and $\gamma$ identification
criteria were the same as used in the analysis of real
data. The $e^+e^-\gamma$ distribution obtained from this simulation is
completely accounted for by combinatorial background from mixed events
(Fig. \ref{fig:pythia_jpsi_gamma_sim}(a)), leaving little room for
contributions from possible jets containing \jpsi, radiative decays or
electron radiation when crossing the detector support.  

Using the data, a check was done for possible missing correlated
radiation backgrounds that might have been missing in the simulation.
The invariant $e^+e^-\gamma$ mass distribution was formed in which we
required $M_{e^+e^-} [\gevcsq]<$2.9.  The \chic contribution is small
in this region and the correlated signal should be mainly from other
sources, e.g. \jpsi internal and external radiation. The data unlike
the simulation shows a correlated background 
after combinatorial background subtraction 
(Fig. \ref{fig:pythia_jpsi_gamma_sim}(b)). The line shape of
this mass distribution can be described by a Gaussian distribution, Landau
distribution, or a simulated $\psip \rightarrow \jpsi+\gamma$ shape.
Its source could be the $gg \rightarrow \jpsi+\gamma+gg$
process mentioned previously, but we simply take this as a background
which must be included in the fit to the \chic mass distribution. The position
of the peak is set by the minimum photon energy cut of 300 MeV, while the width
is set by energy spectrum of the source and more importantly by the 
smearing effect caused by the fact that the spectrum is a difference of
two invariant mass calculations. These results will be used later when
fitting the $\ee\gamma$ invariant mass distribution.

\begin{figure}[!ht]
  \centering
  \includegraphics[width=1.0\linewidth]{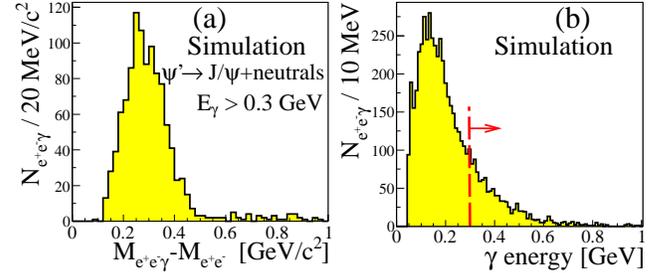}
  \caption{\label{fig:psip_eegamma}(Color online) Mass (a) and
    $\gamma$ energy (b) distributions of $\psip \rightarrow
    \jpsi+$neutrals $\rightarrow \ee\gamma$ obtained from simulations.
    Dashed lines in panel (b) represents the photon energy cut applied
    in this analysis. Appropriate scaling indicates that such events
    contribute between 6 and 20 counts to the correlated $\ee\gamma$
    distribution.}
\end{figure}

In section \ref{sec:psip_jpsi_ratio} we reported that (9.6 $\pm$
2.4)\% of the \jpsi counts in our sample come from \psip
decays. (41.4 $\pm$ 0.9)\% of these decays contain a neutral meson
that decays into photons\cite{Amsler:2008zzb}, namely $\psip
\rightarrow \jpsi + \pi^0\pi^0$, $\psip \rightarrow \jpsi + \pi^0$ and
$\psip \rightarrow \jpsi + \eta$. We will refer to these decay channels
collectively as  $\psip \rightarrow \jpsi + n\gamma$. Simulations show that most of the
decays into neutral mesons are either not detected in the central arm
acceptance or are rejected by the $\gamma$ energy cut, leaving
an estimated 6-20 counts in the low mass distribution of $\ee\gamma$
(Fig. \ref{fig:psip_eegamma}). Contributions from $\psip \rightarrow
\gamma + \chi_c \rightarrow 2\gamma + \jpsi$ decays are expected to be
no larger than three counts.

The contribution from $B$ decays in the $\ee\gamma$ sample was
calculated using the bottom cross section measured by PHENIX
\cite{PhysRevLett.103.082002}. The contribution of $B$ decays to \jpsi
plus at least one photon is less than 3 counts in the entire
$e^+e^-\gamma$ sample.

\begin{figure}
  \centering
  \includegraphics[width=1.0\linewidth]{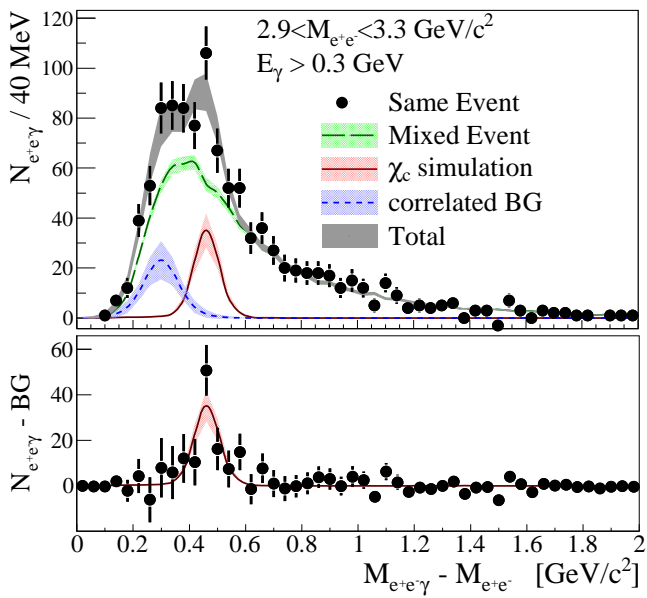}
  \caption{\label{fig:massdist_jpsigamma} (Color online) Top plot:
    $e^+e^-\gamma$ invariant mass distribution where the dielectron
    is required to have an invariant mass within \jpsi mass region.  The
    like-sign dielectron distribution is subtracted as described in
    the text. Bands represent the contributions from various sources:
    solid line - \chic signal; long dashed line - mixed event
    uncorrelated background; short dashed line - correlated background
    as described in the text. The correlated background was fit to a
    Gaussian for this plot.  The widths of the bands reflects the
    1$\sigma$ variations in the fit. The bottom plot is the \chic
    signal after subtraction of the backgrounds.}
\end{figure}

The number of \chic decays was obtained by fitting the background and
the simulated \chic line shapes to the measured $e^+e^-\gamma$ mass
distribution (Fig. \ref{fig:massdist_jpsigamma}). The background
includes two sources: the mixed event background from random \ee +
$\gamma$ combinations, and the correlated background discussed
previously. The correlated background was fit to a Gaussian and a
Landau distribution where the maximum of the
correlated background was set by the photon energy cut.  In addition,
the $\psip \rightarrow \jpsi + n\gamma$ background was used as a third
shape in estimating the systematic error. However, it must be
emphasized that this background cannot explain the magnitude of the
correlated background.  The variations introduced by using the three
distributions contribute to the type B systematic errors.  The fitting
parameters included the combinatorial background normalization, the
amplitude of the correlated background and, when used, the width of the
Gaussian and Landau shapes (the $\psip \rightarrow \jpsi + n\gamma$
shape was fixed from simulations), and the normalization of the
simulated \chic mass distribution. The fitted \chic mass spectrum
returned an average value 96 $\pm$ 24 counts in the \chic mass range
$M_{e^+e^-\gamma}-M_{e^+e^-}\in[0.3,0.6]$ \gevcsq when fitting the
three different line shapes to
the correlated background. 
The signal/background, including the correlated background, was
1/5. The number of \chic counts changed by $\pm$ 4.6\% when using
different line shapes for the correlated background (Gaussian, Landau
or $\psip \rightarrow \jpsi + n\gamma$ shapes).

\subsection{Feed-down fraction result}
\label{sec:chic_result}

The fraction of \jpsi counts coming from \chic decays is
\begin{eqnarray}
  \label{eq:chic_feed_down}
  F_{\chic}^{J/\psi} = \frac{N_{\chi_c}}{N_{J/\psi}}\frac{1}{\mean{\varepsilon_{\chi_c}/\varepsilon_{J/\psi}}}.
\end{eqnarray}
To find the mean conditional acceptance,
$\mean{\varepsilon_{\chi_c}/\varepsilon_{J/\psi}}$, the conditional acceptance
shown in Fig. \ref{fig:chic_eff} must be convoluted with the \chic \pt
distribution.  An estimate of the \chic \pt distribution was obtained
by fitting a two dimensional $e^+e^-\gamma$ mass vs. \pt distribution
to a \chic signal plus backgrounds and extracting the number of \chic
counts in several \pt bins. While the statistical errors are large,
the dependence of the {\it acceptance} on the \pt of the \chic is
mild, hence the error in the mean conditional acceptance is small. We
obtain ($\varepsilon_{\chi_c}/\varepsilon_{J/\psi}$) = \mbox{(12.0
  $\pm$ 0.4) \%}.

Tests of the fitting procedure and the conditional acceptance
calculation were performed using several different simulated data sets
with varying amounts of \chic signal, \jpsi, and  backgrounds. The
feed-down observed after full analysis of the six sets of simulated
events correctly returned the fraction of \chic events with no
significant bias. Variations in the minimum $E_{\gamma}$ criteria
changed the measured feed-down in the simulation by 1.7\%. This
variation is taken into account in the uncertainties as a type B error
introduced by the analysis procedure. 
When the photon energy resolution is changed in a manner consistent
with the measured $\pi^0 \rightarrow 2\gamma$ mass resolution, both
the conditional acceptance and the \chic counts returned from the fits
change, leading to a variation of the feed-down fraction by 1.6\%. The
list of all systematic uncertainties is shown in Table \ref{tab:chic_sys_error}.

\begin{table}[!ht]
  \caption{\label{tab:chic_sys_error}Summary of the type B systematic uncertainties in the \chic
    feed-down fraction measurement. The total gives the sum of all errors in quadrature.}
  \centering
  \begin{ruledtabular}\begin{tabular}{lcc}
    syst uncertainty & contribution & type\\\hline
    $\gamma$ ID & 0.7\% &B \\
    energy resolution & 1.6\% &B \\
    \chic polarization & 1.8\% &B \\
    correlated background line shape & 1.5\% &B \\
    \jpsi continuum & 0.1\% &B \\
    fit procedure & 1.7\% &B \\
    \chic momentum dependence & 1.1\% &B \\ \\
    TOTAL & 3.6 \%\\
  \end{tabular}\end{ruledtabular}
\end{table}

The final \chic feed-down fraction using (\ref{eq:chic_feed_down}) is
\begin{eqnarray}
  \label{eq:chic_feeddown}
  F_{\chi_c}^{J/\psi} = 32 \pm 9 \%
\end{eqnarray}
\noindent when taking the quadratic sum of the statistical and
systematic uncertainties.

\section{\jpsi analysis in the forward rapidity region}
\label{sec:dimuon_analysis}

This section describes the analysis performed to obtain the inclusive
\jpsi dimuon yield at forward rapidity $1.2<|y|<2.4$. Section
\ref{sec:dimuon_signal_extraction} describes the \jpsi signal
extraction from the dimuon spectrum and related uncertainties. The
response of the muon arm spectrometers to dimuon decays from the
\jpsi is described in section \ref{sec:dimuon_acceptance}. Finally, the
\pt and rapidity dependence of the \jpsi differential cross section
and a summary of systematic uncertainties is reported in section
\ref{sec:dimuon_cross_section}.

\subsection{$\jpsi \rightarrow \mu^+\mu^-$ signal extraction}
\label{sec:dimuon_signal_extraction}

The dimuon invariant mass spectrum was obtained from the muon sample selected
according to the criteria described in Sec. \ref{sec:apparatus}. The
MuID trigger condition is emulated offline. In order to make sure the
real \jpsi candidate fired the MuID trigger, at least one muon of the
dimuon pair is required to match a road from the trigger emulator.

\begin{figure}
  \centering
  \includegraphics[width=1.0\linewidth]{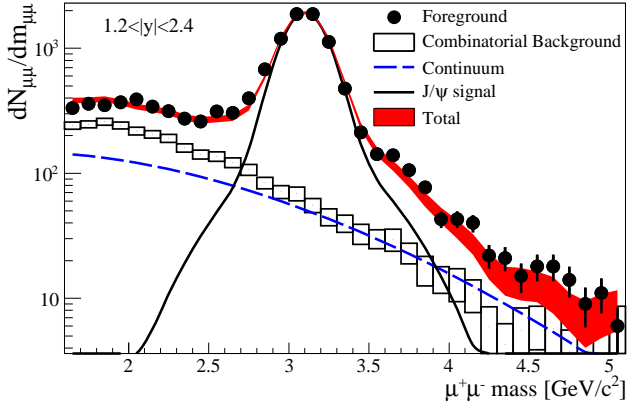}
  \caption{\label{fig:jpsi_peak_forward}(Color online) Invariant mass
    distribution of dimuons in the \jpsi mass region.  The components
    of the spectrum are the combinatorial background estimated from a
    mixed-event technique, an acceptance modified
    (Fig. \ref{fig:acc_dimuon_mass}) two-Gaussian \jpsi signal, and an
    acceptance modified exponential continuum
    (Eq. \ref{eq:dimuon_fit}).}
\end{figure}

The decomposition of the dimuon background is very similar to that
described in Sec. \ref{sec:dielectron_id} for dielectrons. The
combinatorial background was estimated using the mass spectrum of
random pairs formed by pairing opposite sign muon candidates from
different events. The muons of the mixed pair are required to have
vertices that differ by no more than 3 cm in the beam direction. The
mixed event spectrum was normalized by the factor
\begin{eqnarray}
\alpha=\frac{\sqrt{(N_{\mu^+\mu^+}^{\rm same})(N_{\mu^-\mu^-}^{\rm same})}}{\sqrt{(N_{\mu^+\mu^+}^{mixed})(N_{\mu^-\mu^-}^{mixed})}},
\end{eqnarray}

\noindent where $N_{\mu\mu}^{\rm same}$ and $N_{\mu\mu}^{mixed}$ are the
number of pairs formed from two muons in the same or in mixed events, 
respectively.  The mass spectrum of the dimuons in the \jpsi mass region
is shown in Fig. \ref{fig:jpsi_peak_forward}.

\begin{figure}[!ht]
  \centering
  \includegraphics[width=1.0\linewidth]{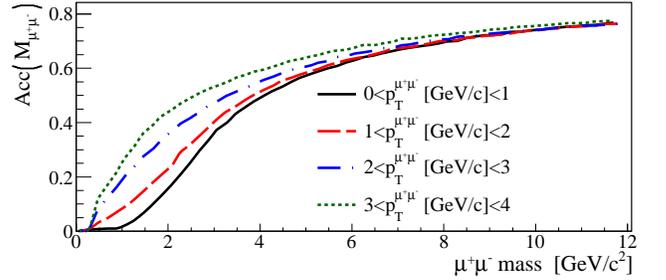}
  \caption{\label{fig:acc_dimuon_mass}(Color online) Mass dependence
    of the dimuon geometric acceptance in the muon arm
    spectrometers.}
\end{figure}

The components of the correlated dimuon spectrum are muons
sharing the same $c\bar{c}$ or $b\bar{b}$ ancestor, dimuons from
Drell-Yan and the \jpsi and \psip resonances. There is no clean
mass discrimination between the \jpsi and \psip mass peaks in the muon arm
spectrometers. However the \psip contribution is expected to be negligible
in the peak integral compared to other uncertainties. The correlated
dimuon mass distribution can be represented by a function
$F(M_{\mu\mu})$ including an exponential shape accounting for the
continuum distribution, a double Gaussian which describes the line
shape of \jpsi in the Monte Carlo, and acceptance dependence:

\begin{eqnarray}
  \label{eq:dimuon_fit}
 \centering
  \frac{F(M_{\mu\mu})}{Acc\left(M_{\mu\mu}\right)} &=& 
  A_{\psi}F_{\psi}(M_{\mu\mu}) + A_{cont}e^{-\frac{M_{\mu\mu}}{b_{cont}}}\\\nonumber
  F_{\psi}(M_{\mu\mu}) &=& (1-f_{G2})G\func{M_{\mu\mu},M_{J/\psi},\sigma_{G1}} \\\nonumber
  &+& f_{G2}G\func{M_{\mu\mu},M_{J/\psi}+\delta_{M},\sigma_{G2}} \\\nonumber
  G\func{M_{\mu\mu},M,\sigma} &=& \frac{1}{\sqrt{2\pi}\sigma} e^{-\frac{(M_{\mu\mu}-M)^2}{2\sigma^2}}\\\nonumber
\end{eqnarray}

\noindent where $Acc\left(M_{\mu\mu}\right)$ is the mass
dependence of the dimuon acceptance in the rapidity
1.2$<|y|<$2.4 estimated using dimuon simulation
(Fig. \ref{fig:acc_dimuon_mass}), $A_{\psi}$ is the amplitude of the
\jpsi signal with mass $M_{J/\psi}$ composed of a Gaussian of width
$\sigma_{G1}$ and a second Gaussian of width $\sigma_{G2}$ shifted by 
$\delta M$ in mass. $f_{G2}$ describes the fractional strength of the
second Gaussian.  The normalization of the continuum contribution is
$A_{cont}$ and its exponential slope is $b_{cont}^{-1}$.

The correlated mass distribution function $F\left(M_{\mu\mu}\right)$
was fit to the measured unlike-sign dimuon mass distribution for
each \pt and rapidity range using the maximum likelihood method.
The combinatorial background, obtained from the normalized mixed event
distribution, was also introduced in the fit with a fixed amplitude.
The mass resolution obtained in the entire \jpsi sample was
$\sigma_{G1}/M_{J/\psi}=$4\% ($\sigma_{G1} = 125$ MeV). The fitting
parameters which determine the 
line shape of the \jpsi peak ($\sigma_{G1}$, $\sigma_{G2}$,
$\delta_M$ and $f_{G2}$) obtained from the entire unbinned sample
were fixed when performing fits for individual \pt and
rapidity bins. The \jpsi mass, $M_{J/\psi}$, was allowed to vary by
10\% of its nominal value \mbox{(3.096 \gevcsq)} in the fitting procedure,
the \jpsi and continuum amplitudes were constrained to avoid
unphysical negative values, and the exponential slope was allowed to
vary by 20\% from a the value found in a fit to the entire (unbinned)
sample. For the systematic uncertainty 
evaluation, $f_{G2}$ was changed by 25\% up and down, the fit was
performed in two mass ranges: 1.8 $<M_{\mu\mu} [\gevcsq]<$7.0 and
2.2$<M_{\mu\mu} [\gevcsq]<$6.0 and the combinatorial background
normalization $\alpha$ was varied by $\pm$2\%. Fig.
\ref{fig:jpsi_peak_forward} shows the fitted function and its
components for the dimuon unlike-sign distribution for one of the
rapidity bins. Two methods for counting the {\jpsi}s were considered: 1) using the
fitted amplitude $A_{\psi}$ directly, or 2) from direct counting of dimuon pairs in the mass
region 2.6\ $<M_{\mu^+\mu^-} [\gevcsq]<$\ 3.6 with subtraction of the combinatorial
background and exponential continuum underneath the peak in that same region. The standard 
deviations of the central values of the fits and of the signal extraction method
variations are taken as type A signal extraction systematic
uncertainties, since these variations are largely driven by statistical variations.
The total number of \jpsi counts was 
$16,612 \pm 147^{\rm stat} \pm 112^{\rm syst}$ 
in the south muon arm and 
$16,669 \pm 145^{\rm stat} \pm 115^{\rm syst}$ 
in the north muon arm.

\subsection{Di-muon acceptance and efficiency studies}
\label{sec:dimuon_acceptance}

\begin{figure}
  \centering
  \includegraphics[width=1.0\linewidth]{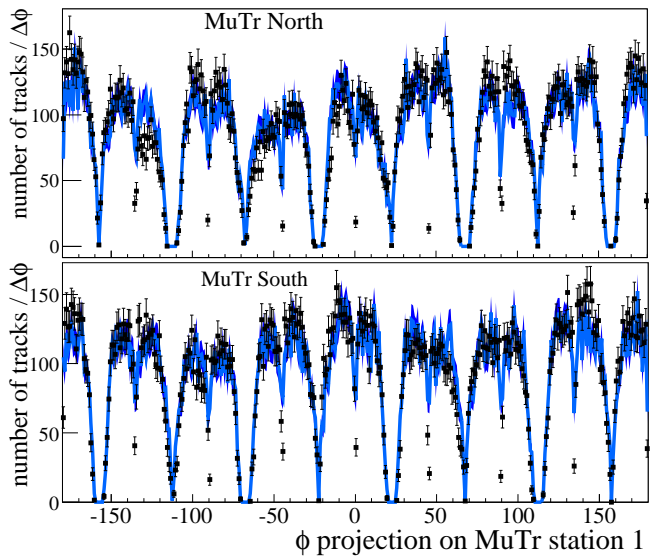}
  \caption{\label{fig:acc_phi_muon}(Color online) Simulated (shaded
    area) and real data (bars) single muons
    distributed in the $\phi$ coordinate of the MuTr.}
\end{figure}

The response of the muon arm spectrometers to dimuons from {\jpsi} decays was 
studied using a tuned {\sc geant}3-based simulation of the muon arms and an offline MuID
trigger emulator. The MuID panel-by-panel efficiency used in these
simulations was estimated from reconstructed roads in real data, or in
cases with low statistics, from a calculation based on the operational
history record for each channel. The MuID efficiency had a variation
of 2\% throughout the Run leading to a systematic uncertainty of 4\%
for the \jpsi yield.

The charge distribution in each part of the MuTr observed in real data and
the dead channels and their variation with time over the run were used
to give an accurate description of the MuTr performance within the detector simulation. 
The azimuthal distribution of muon candidates in real data and simulated
muons from \jpsi decays using the {\sc pythia} simulation are shown in Fig.
\ref{fig:acc_phi_muon}. The $Z$ vertex distribution of simulated \jpsi
decays is the same as that observed in real data. The \pt distribution
in the MuTr obtained in simulation was also weighted according to
that observed in the real data. The relatively small differences in the
real and simulated $\phi$ distributions (Fig. \ref{fig:acc_phi_muon}) are thought to be due
primarily to missing records for short periods of time in the dead HV channel records.
These differences are estimated to change the \jpsi dimuon yields by
up to 6.4(4.0)\% in north(south) arms. Run-by-run variations of the MuTr
single muon yields are estimated to affect the final \jpsi yields by an additional 2\%.

\begin{figure} 
\centering
  \includegraphics[width=1.0\linewidth]{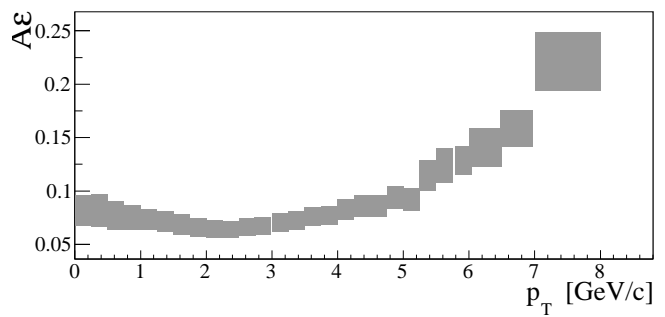}
  \caption{\label{fig:jpsi_dimuon_efficiency} Transverse momentum dependence
    of the north and south average muon arms acceptance $\times$
    efficiency for \jpsi dimuon decays  in $1.2<|y|<2.4$. Shaded
    bands are the uncertainties of the estimates described in the text.}
\end{figure}

The \jpsi acceptance $\times$ efficiency ($A\varepsilon$) evaluation
used a {\sc pythia} simulation with several parton distributions as input to
account for the unknown true rapidity dependence of the \jpsi
yield leading to variations of 4\% in the final
acceptance. Fig. \ref{fig:jpsi_dimuon_efficiency} shows the overall
\pt dependence of $A\varepsilon$ for \jpsi dimuon decays.
The uncertainties related to the
knowledge of the detector performance are point-to-point correlated between different
\pt and different rapidity bins. The uncertainty in the dimuon acceptance
caused by lack of knowledge of the \jpsi polarization was studied using the detector
simulation. The first results in PHENIX at forward rapidity \cite{PPG121}
indicate that the \jpsi polarization is no larger than $\pm$0.5 for $p_T<$5
\gevc (in the Helicity frame). 
For this polarization variation, the simulations show one standard deviation
variations between 2\% and 11\%, with the largest variation occurring 
for $p_T<$1 \gevc and $y \simeq$ 1.2.
For $p_T>5$ \gevc, where there are no polarization measurements
we consider polarizations anywhere between $\pm$1, and find variations
no larger than 5\%. These deviations are considered as type B uncertainties.

\subsection{\jpsi dimuon cross section result}
\label{sec:dimuon_cross_section}

The differential cross section for each \pt bin was calculated
according to Eq. (\ref{eq:cross_section}). The systematic uncertainties involved
in this calculation are listed in Table \ref{tab:sys_error_dimuon}.

\begin{table}[!ht]
  \caption{\label{tab:sys_error_dimuon} List of the systematic uncertainties
    in the \jpsi dimuon yield measurement. Ranges indicate \pt dependence.}
  \begin{ruledtabular}\begin{tabular}{lcc}
    description & relative uncertainty & type\\\hline
    signal extraction & 1.8\% - 35\% & A \\
    MuID efficiency & 4\% & B \\
    MuTr acceptance & 6.4\%(north), 4.0\%(south) & B \\
    run-by-run fluctuation & 2\% & B \\
    Monte Carlo \jpsi input & 4\% & B \\ 
    \jpsi polarization & 2\% - 11\% & B \\
    luminosity & 10\% & C\\
\end{tabular}\end{ruledtabular}
\end{table}

\begin{figure} 
\centering
  \includegraphics[width=1.0\linewidth]{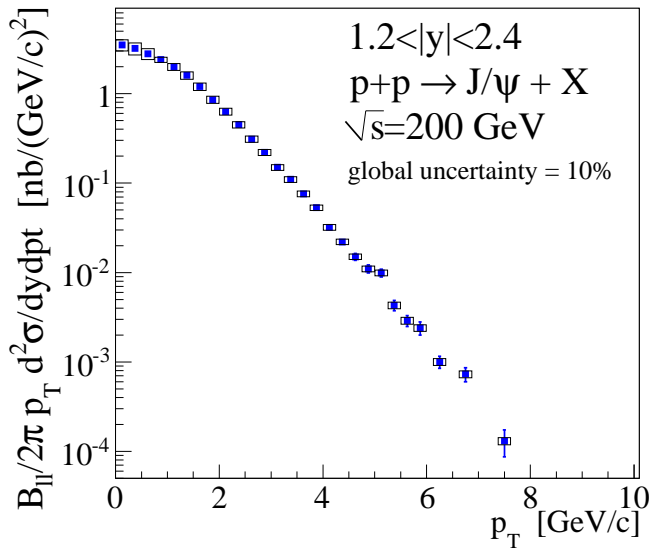}
  \caption{\label{fig:jpsi_dimuon} Transverse momentum dependence
    of the \jpsi dimuon differential cross section obtained in the
    muon arms in 2006 and 2008 Runs.}
\end{figure}

The differential cross section was independently
obtained in the north and south muon arm spectrometers and for the 2006 and
2008 Runs. The measurements agree in all data sets for all \pt points
within one sigma statistical and systematic uncertainties.
The averaging of these four momentum spectra is done using a weight
for each data set based on the uncertainties for each that are uncorrelated between data sets.
By definition the statistical and type A uncertainties are
uncorrelated and while the type C is correlated. The uncertainties in the MuTr
efficiency and run-by-run variations are also uncorrelated between
data sets. The MuID efficiency and the simulation input uncertainty
are correlated between different spectrometer arms and run periods. Fig.
\ref{fig:jpsi_dimuon} shows the resulting average differential cross section for
dimuons from \jpsi. The numbers are listed in Table
\ref{tab:jpsi_yield_forward}. 

The rapidity distribution was calculated as:

\begin{eqnarray}
  B_{\mu\mu}\frac{d\sigma}{dy} = \frac{N_{J/\psi}}{\Delta y
    \varepsilon_{inel.} \int \mathcal{L}dt A\varepsilon},
\end{eqnarray}

\noindent where the number of \jpsi counts $N_{J/\psi}$ and the
acceptance $\times$ efficiency estimates were performed for each
rapidity bin.
All these results are shown in Fig.
\ref{fig:jpsi_rap} and the numerical results are listed in Table
\ref{tab:jpsi_rap}.

\section{Results Discussion}
\label{sec:results_discussion}

This section presents a summary of the results reported in the
previous sections and compares them with results obtained in other
experiments as well as predictions from several different production
mechanism calculations.  The rapidity dependence of the \jpsi yield is
compared to models using various parton distribution functions (PDF)
in Sec. \ref{sec:jpsi_totcross_sec}.  The total \jpsi
cross section is derived from the rapidity distribution and discussed
in Sec. \ref{sec:total_xsec}.  The \jpsi differential cross section
dependence on \pt is compared to empirical scaling laws observed at
lower energies as well as different charmonium hadronization models in
Sec. \ref{sec:jpsi_pt}.  The measured fraction of the \jpsi yield
coming from \psip and \chic decays is compared to other experiments in
Sec. \ref{sec:feed-down}. The consequences of the results presented in
this article on recent charmonium measurements in $p$(d)+A and A+A
collisions is the subject of the Sec. \ref{sec:outcomes}.

The models used in our comparisons were described in
Sec. \ref{section:introduction}; namely, the Color Evaporation Model
(CEM), the Color Singlet Model (CSM) and Non-relativistic QCD (NRQCD).
The CEM used FONLL calculations for the charm cross section and CTEQ6M
as the parton distribution function\cite{Frawley:2008kk,Ramona:2009}.
For the CSM comparison, we used the recent NLO calculation only for the
direct \jpsi yield at RHIC energy and PHENIX rapidity coverage
\cite{Lansberg:2010vq}.  We used two NRQCD calculations in our
comparisons. The calculation performed for the direct \jpsi plus \chic
feed-down in \cite{Butenschoen:2010rq} uses NLO diagrams for the color
singlet and color octet states with a long range matrix element tuned from
experimental hadroproduction \cite{PhysRevD.71.032001} and photoproduction
\cite{Adloff:2002ex,Aaron:2010gz} results. This
calculation is only available for the differential \pt dependent cross
section. An older calculation, performed for the same direct \jpsi plus
\chic feed-down with LO diagrams \cite{Cooper:2004qe}, also provides
the rapidity dependence and total cross sections for different
PDFs. No similar attempt has been made with the new calculations.
The differential \pt
dependent cross section calculation involves the emission of a hard gluon 
which determines the shape of the charmonium \pt spectrum. The amplitude of the 
hard gluon emission cannot be calculated for $p_T<2$ \gevc because of infrared 
divergences. This problem is circumvented in the older calculation
by empirically constraining the low \pt  nonperturbative soft gluon emission
to obtain the rapidity dependence, $d\sigma/dy$. 
In both NRQCD calculations there is a prevalence of color octet
states in the direct \jpsi contribution.

\subsection{\jpsi Rapidity dependence}
\label{sec:jpsi_totcross_sec}

\begin{figure}
  \centering
  \includegraphics[width=1.0\linewidth]{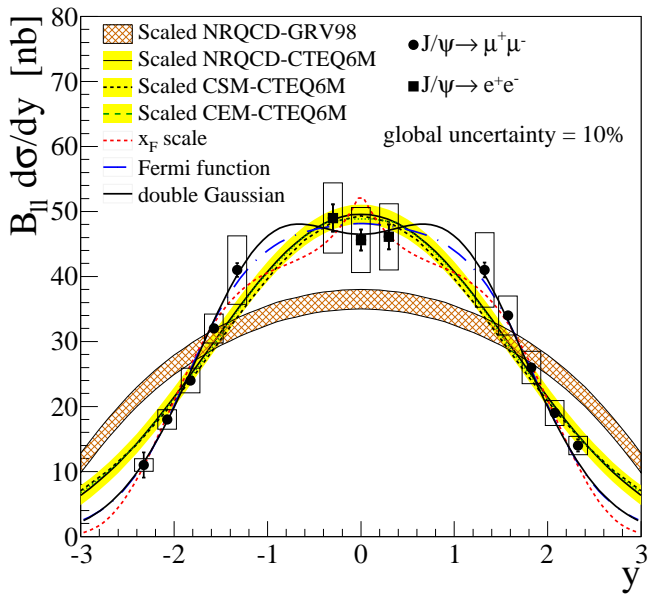}
  \caption{\label{fig:jpsi_rap} (Color online)Rapidity dependence of
    the \jpsi yield combining dielectron ($|y|<0.35$ - full squares)
    with dimuon channels ($1.2<|y|<2.4$ - full circles) along with
    the fits used to estimate the total cross section.  Lines
    correspond to the three fitting functions described in the text.
    Also shown are arbitrarily normalized model predictions (NRQCD
    \cite{Cooper:2004qe}, CEM \cite{Frawley:2008kk,Ramona:2009} and
    CSM \cite{Lansberg:2010vq}).}
\end{figure}

The rapidity distribution of the \jpsi dilepton cross section is
shown in Fig. \ref{fig:jpsi_rap} and in Table \ref{tab:jpsi_rap}.
The data points are grouped into three rapidity ranges, corresponding
to the different detectors used in the measurement: 
south muon arm ($-2.4<y<-1.2$), central arms ($|y|<0.35$) and north
muon arm ($1.2<y<2.4$). The systematic uncertainties represented by
the boxes are point-to-point correlated for data points in the same
group and are uncorrelated between different groups. All points have a
global uncertainty of 10\% coming from the minimum-bias trigger
efficiency estimate.

In order to compare the shape of the rapidity distribution, we
normalized the CEM, CSM and NRQCD predictions to the
integral of the measured data in Fig.\ref{fig:jpsi_rap}. All models
use the CTEQ6M PDFs. The NRQCD model is also available with
the GRV98 and the MRST99 PDFs. The theoretical
rapidity distributions exhibit a similar shape when using CTEQ6M. A
very different rapidity distribution is obtained when the NRQCD prediction is
calculated using GRV98 and MRST99 (MRST99 is not shown in the
figure). These observations suggest that the choice of PDF plays
the most important role in describing the shape of the \jpsi rapidity
distribution. 
The rapidity shape also appears to be independent of the feed-down contributions,
since the CSM has a similar shape to the CEM and NRQCD model, despite the fact that it 
contains only direct \jpsi contributions. The PDF which
best describes the data is CTEQ6M and we use this for the remaining
comparisons. 

An empirical description of the \jpsi yield used in some
fixed-target experiments with large coverage is based on
the Feynman $x_F$ form \cite{Schub:1995pu},

\begin{eqnarray}
  \frac{d\sigma}{dx_F} &=& A\left(1-|x_F|\right)^c.
\end{eqnarray}

\begin{figure}
  \centering
  \includegraphics[width=1.0\linewidth]{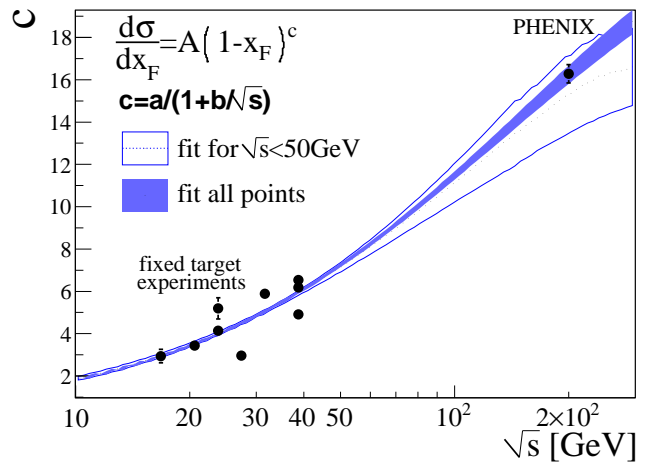}
  \caption{\label{fig:cpar_sqrts} (Color online) Energy dependence of
    the parameter $c$ where $d\sigma/dx_F = (1-x_F)^c$ is fitted to $x_F$
    distributions of \jpsi production in fixed target experiments
    \cite{Anderson:1976ja,Branson:1977ci,Binkley:1976rn,Antoniazzi:1992af,Siskind:1979ax,Gribushin:1999ha,Schub:1995pu,Alexopoulos:1997yd}
    and in the PHENIX rapidity distribution. The parameters
    ($a=27.8 \pm 1.5$, $b=(143 \pm 11)$ GeV) are obtained from a fit to the
    experimental data points.}
\end{figure}

We can convert to a rapidity distribution by writing
\begin{eqnarray}
  \label{eq:feynmann_func}
  \frac{d\sigma}{dy} &=& \frac{d\sigma}{dx_F} \frac{dx_F}{dy} \\\nonumber
  &\approx& 2A\left(1-|x_F|\right)^c \sqrt{
\frac{\mean{p_T^2}+M_{J/\psi}^2}{s}
}\cosh{y}
\end{eqnarray}

\noindent where $p_T = 1.73$ \gevc is the average of the \jpsi \pt
distributions over all measured rapidities. The fit returned $c=16.3
\pm 0.4$ with $\chi^2$ probability of 31\%, where statistical and
systematic uncertainties are summed in quadrature(Fig. \ref{fig:jpsi_rap}). Fig.
\ref{fig:cpar_sqrts} shows that $c$ scales approximately as
$c=a/\left(1+b/\sqrt{s}\right)$. This extrapolation of the rapidity
dependence can be used to estimate the total cross section from
measurements with limited rapidity coverage and 
will be used as one method to calculate the total cross section
from the present measurement.

\subsection{Total cross section of inclusive \jpsi}
\label{sec:total_xsec}

The total cross section was estimated from different empirical
functions fitted to the rapidity distribution - a double
Gaussian,the $x_F$ scaling function (\ref{eq:feynmann_func}) described above, and a ``Fermi'' function:
\begin{eqnarray}
  \label{eq:fermi_func}
\frac{d\sigma}{dy} = \frac{A}{1 + e^{\frac{(y-\lambda)}{\sigma}}}
\end{eqnarray}

\noindent Rapidity distributions based on charmonium production models were not
used in the total cross section in order to avoid any theoretical
bias.

The correlated uncertainties between data points measured in 
each spectrometer were propagated to the fit uncertainty by 
allowing the points to move coherently in the rapidity range 
covered by that spectrometer.  Table \ref{tab:rap_fit} shows 
the total dilepton cross section and the $\chi^2$ probability 
for each function used in the fit. The final cross section is 
obtained from the average of the numbers from each fit function 
weighted according to their $\chi^2$ probability.  The 
systematic uncertainty from the unknown rapidity shape is taken 
from the standard deviation between the three fitting 
functions.  Based on these fits, we conclude that the PHENIX 
rapidity acceptance covers \mbox{56 $\pm$ 2 \%} of the total 
\jpsi cross section. The \jpsi cross section reported in this 
paper is $180.7 \pm 2.0^{\rm stat} \pm 11^{\rm syst} nb$, in 
agreement with our previous result with a reduction in the 
statistical and systematic uncertainties.

\begin{table}[!ht]
  \centering
  \caption{\label{tab:rap_fit} Estimate of the total dilepton \jpsi
    cross section from the three fitting functions, together with the weighted
    average and a comparison to the result obtained in our previous measurement.
    The measured total cross sections have an additional 10\% global uncertainty.
\\}
  \begin{ruledtabular}\begin{tabular}{lcl}
    estimating function & $\chi^2$ prob.  & 
    \ \ \ \ \ \ \ \ \ $B_{ll}\sigma_{\jpsi} (nb)$ \\ \hline

    $x_F$ scale fcn, Eq.\ref{eq:feynmann_func} & 0.30 & 
    $170.8 \pm 1.5^{\rm stat} \pm 9^{\rm syst}$ \\

    double Gaussian & 0.79 & 
    $183.5 \pm 1.9^{\rm stat} \pm 11^{\rm syst}$ \\

    Fermi fcn, Eq.\ref{eq:fermi_func} & 0.70 & 
    $182.0 \pm 2.3^{\rm stat} \pm 12^{\rm syst}$ \\ \\

    AVERAGE & & 
    $180.7 \pm 2.0^{\rm stat} \pm 12^{\rm syst}$ \\ \\

    2005 Run result\cite{Adare:2006kf} & &
    $178 \pm 3.0^{\rm stat} \pm 53^{\rm syst}$ \\
  \end{tabular}\end{ruledtabular}
\end{table}


Table \ref{tab:tot_Xsec} presents the measured total \jpsi cross
section and the expectations from the three production models
considered in this text. The experimental direct \jpsi cross section
is estimated assuming that the feed-down fraction of \chic and \psip
measured at midrapidity is the same at forward rapidity. The
feed-down from $B$ mesons is only significant at high \pt and is not
considered in the total cross section.  The total cross section
estimated using the CEM is the only one which agrees with the
experimental result, although the cross section calculation includes
the scale factor $\mathcal{F}$ (Sec. \ref{section:introduction})
obtained from \jpsi measurements.  The NRQCD includes color singlet
and color octet states, and as mentioned at the beginning of this
section, cannot be extrapolated to low \pt to obtain the rapidity
distribution without the addition of an empirical constraint.

\begin{table}[!ht]
  \centering
  \caption{\label{tab:tot_Xsec} Comparison of the measured \jpsi cross
    section with the three models considered in this text. Direct \jpsi 
    cross sections are obtained
    assuming that the \chic and \psip feed-down fractions measured at
    midrapidity are the same at forward rapidity. Type A, type B and type C errors 
    are quadratically summed in the measured result.}
  \begin{ruledtabular}\begin{tabular}{lll}
     & direct \jpsi & inclusive \\\hline
   CEM &\ \ \ \ \ \ - & 169 $\pm$ 30 nb \\
   NLO CSM & 53 $\pm$ 26 nb &\ \ \ \ \ \ - \\
   LO NRQCD &\ \ \ \ \ \ - & 140 $\pm$ 5 nb \\
   Measured & 105 $\pm$ 26 nb & 181 $\pm$ 22 nb\\
  \end{tabular}\end{ruledtabular}
\end{table}

\subsection{\jpsi \pt distribution}
\label{sec:jpsi_pt}

\begin{figure*}
  \centering
  \includegraphics[width=0.75\linewidth]{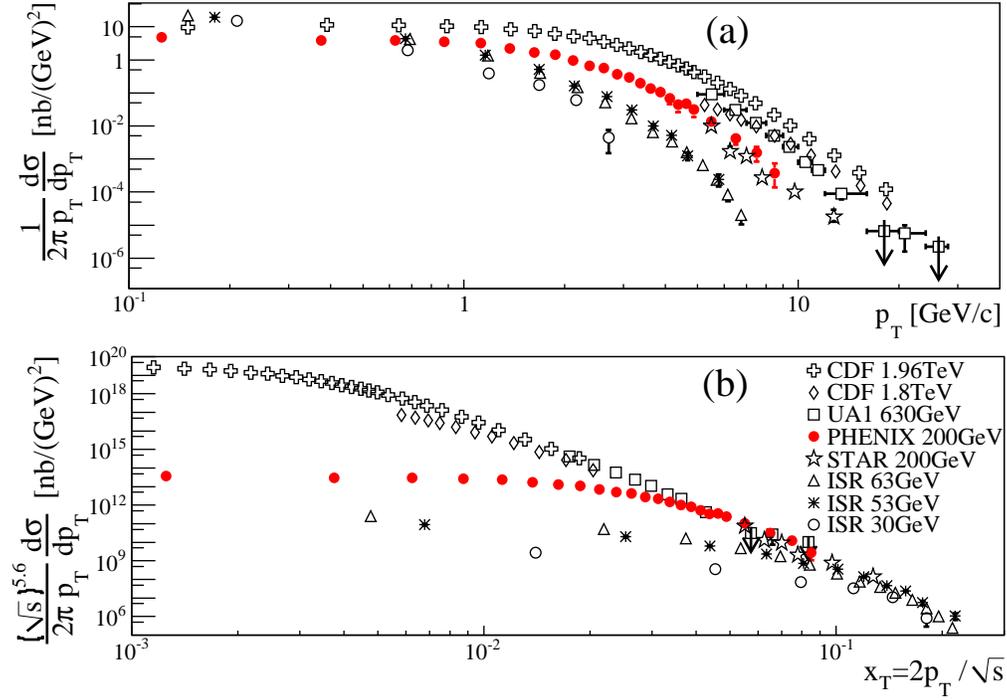}
  \caption{\label{fig:jpsi_pt_xt} (Color online) \pt distribution (a)
    and $x_T$ distribution (b) of \jpsi yield in PHENIX, STAR
    \cite{Abelev:2009qaa}, ISR \cite{Clark:1978mg}, UA1
    \cite{Albajar:1990hf} and CDF\cite{Abe:1997jz,PhysRevD.71.032001}
    at $y \sim 0$.}
\end{figure*}

The \pt-dependent dielectron differential cross section at
midrapidity is compared to other \pp and $p+\bar{p}$ experiments in
Fig. \ref{fig:jpsi_pt_xt}(a). The shapes of the transverse momentum
distributions follow the well known "thermal" exponential behavior for
$p_T<2$ \gevc and a hard-scattering power law behavior at high \pt. The hard
process scales with $x_T=2p_T/\sqrt{s}$ ($\sqrt{s}^{n}
Ed^3\sigma/d^3p = G(x_T)$) \cite{Adcox:2004mh} for all collision
energies, as can be seen in Fig. \ref{fig:jpsi_pt_xt}(b), where
$n=5.6 \pm 0.2$ \cite{Abelev:2009qaa}. n is related to the number of
partons involved in the interaction. A pure LO process leads to
$n=4$, hence, NLO terms may be important in \jpsi production.\cite{Berman:1971xz,Blankenbecler:1975ct,Blankenbecler:1972cd,Cahalan:1974tp}

\begin{figure}
  \centering
  \includegraphics[width=1.0\linewidth]{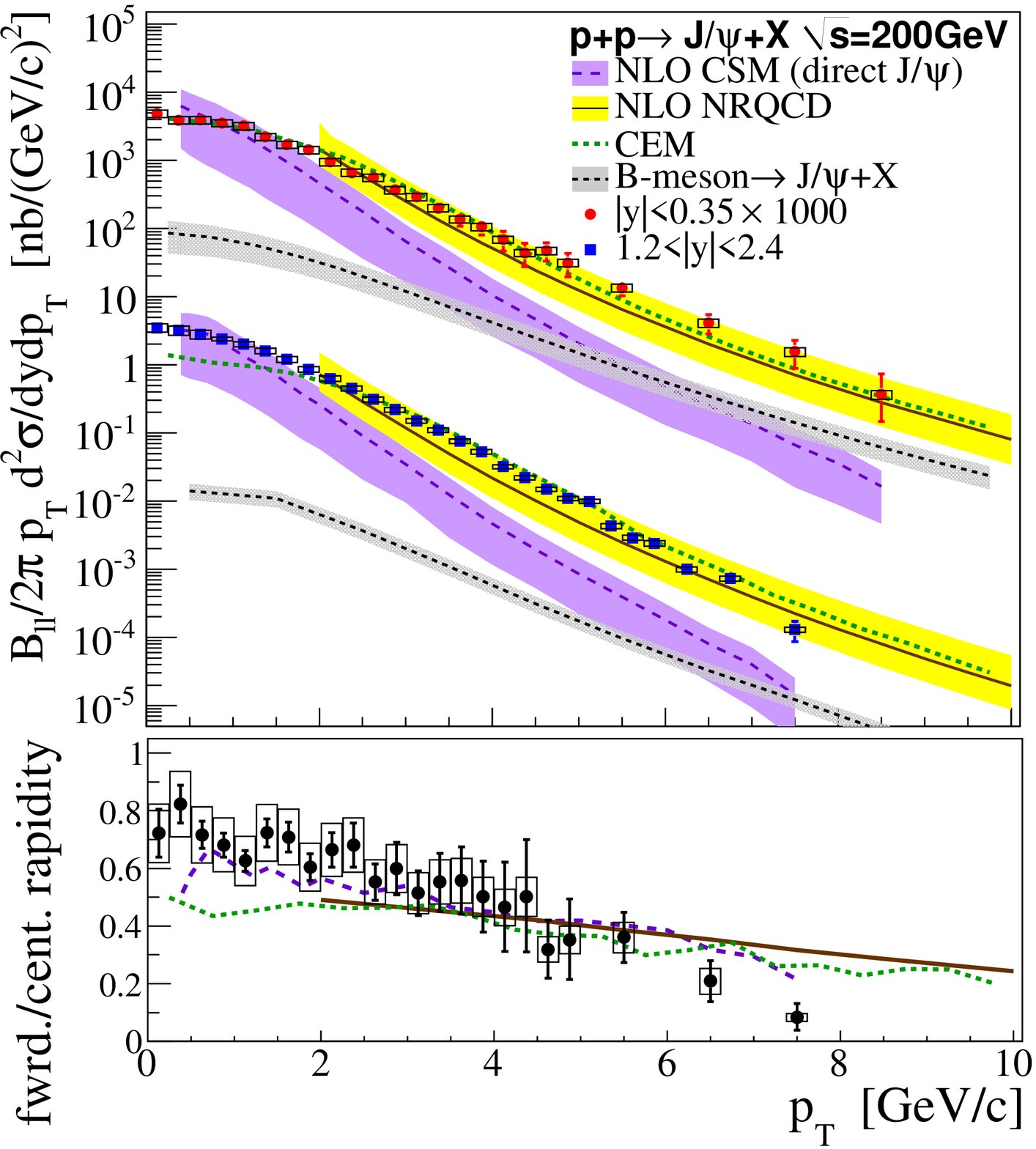}
  \caption{\label{fig:jpsi_central_forward} (Color online) Top:
    Transverse momentum dependence of \jpsi yield in $|y|<0.35$ and
    $1.2<|y|<2.4$ along with predictions based on CSM
    \cite{Lansberg:2010vq}, NRQCD \cite{Butenschoen:2010rq}, CEM
    \cite{Ramona:2009,Frawley:2008kk} and $B$-meson decay based on
    FONLL calculation \cite{PhysRevLett.95.122001}. All models use
    CTEQ6M. Theoretical uncertainties are represented as bands. Note
    that the midrapidity points are scaled up by a factor of 1000.
    Bottom: Ratio of the forward and central rapidity \pt spectra and
    corresponding theoretical predictions.}
\end{figure}

The \pt dependence of the \jpsi differential cross sections measured
at forward and midrapidity are shown in Fig.
\ref{fig:jpsi_central_forward} along with theoretical calculations
where the absolute normalization is determined in the
calculations. 
%
The CEM and the NRQCD (for $p_T>2$ \gevc) provide reasonable descriptions of 
the \pt distribution, whereas the CSM disagrees in both the normalization and 
the slope of the \pt distribution, indicating that NLO color singlet 
intermediate states cannot account for the direct \jpsi production. However, 
the NLO CSM calculation gives a good description of the \jpsi polarization 
measured by PHENIX \cite{PhysRevD.82.012001,Lansberg:2010vq}. Attempts are
being made to extend the CSM to NNLO. 
Preliminary NNLO CSM calculations performed for $p_T>5$ \gevc 
\cite{Lansberg:2010vq} shows a large increase in the yield, but still under-
predict the experimental results.
None of these theoretical models consider the $B$-meson decay
contribution to the \jpsi yield. The fixed-order plus
next-to-leading-log (FONLL) \cite{PhysRevLett.95.122001} calculation
of these decays is also plotted in Fig. \ref{fig:jpsi_central_forward}
and has a reasonable agreement with STAR measurements using
\jpsi-hadron correlations \cite{Abelev:2009qaa}. According to this
calculation, the $B$-meson contribution to the measured \jpsi
inclusive yield is between 2\%(1\%) at 1 \gevc and 20\%(15\%) at 7.5
\gevc in the mid(forward)-rapidity region with large theoretical
uncertainties. 

The $p_{T}$ dependence of the \jpsi yield is harder at midrapidity,
as seen from the ratio between the forward and midrapidity
differential cross sections versus $p_T$ shown in Fig.
\ref{fig:jpsi_central_forward}(bottom).  
The figure also includes the forward/midrapidity
yield ratios from the theoretical models using their mean values and
assuming that theoretical uncertainties in these ratios cancel
out. All of the models predict a downward trend, but the CEM and NRQCD
calculations do not follow a slope as large as the data.

The mean transverse momentum squared \mean{p_T^2} was calculated
numerically from the \pt distribution. The correlated uncertainty was
propagated to \mean{p_T^2} by moving low-\pt and high-\pt data points
coherently in opposite directions according to their type B
uncertainty. The results with the propagated type A and type B
uncertainties are listed in Table \ref{tab:pt2}. The table also
contains \mean{p_T^2} for $p_T<$ 5 \gevc for a direct comparison with
previous PHENIX results \cite{PhysRevLett.101.122301}. As expected,
the mean transverse momentum squared at midrapidity is larger than at
forward rapidity.

\begin{table}[!ht]
  \centering
  \caption{\label{tab:pt2}Mean transverse momentum squared in
    $(\gevc)^2$ of \jpsi and \psip for different rapidity and \pt
    ranges. Uncertainties are type A and type B respectively.
\\}
  \begin{ruledtabular}\begin{tabular}{lcc}
    system & $\mean{p_T^2}$  & $\left.\mean{p_T^2}\right|_{p_T<5GeV/c}$ \\\hline
    \jpsi  $1.2<|y|<2.4$&\  3.65$\pm$0.03$\pm$0.09&\ 3.45$\pm$0.03$\pm$0.08\\
    \jpsi  $|y|<0.35$ &\  4.41$\pm$0.14$\pm$0.11&\ 3.89$\pm$0.11$\pm$0.09\\
    \psip  $|y|<0.35$ &  4.7 $^{+1.5}_{-1.05}$ $ \pm$0.2&4.7$^{+1.5}_{-1.05}$ $\pm$0.2
  \end{tabular}\end{ruledtabular}
\end{table}



\subsection{Charmonia ratios and \jpsi feed-down fractions}
\label{sec:feed-down}

\begin{figure}
  \centering
  \includegraphics[width=1.0\linewidth]{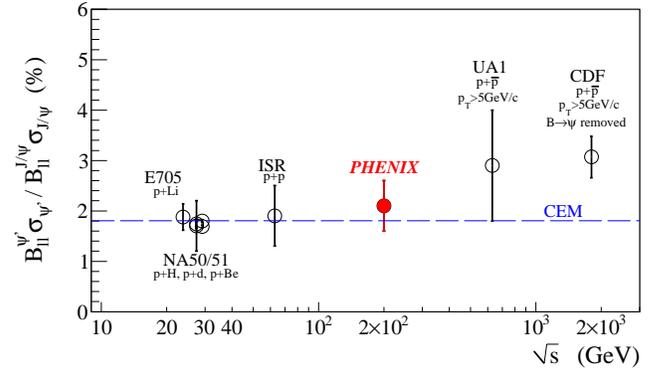}
  \caption{\label{fig:psip_jpsi_sqrts} (Color online) Collision energy
    dependence of the $\psip / (J/\psi)$ dilepton cross section ratio
    obtained in \pp and light fixed target $p$+A collisions 
    \cite{Antoniazzi:1992iv,Abreu:1998rx,Alessandro:2006jt,Clark:1978mg,Albajar:1990hf,Abe:1997jz}
    compared to the CEM estimate \cite{Frawley:2008kk}. Statistical
  and systematic errors were quadratically summed.}
\end{figure}

The transverse momentum dependence of the $\psip/(\jpsi)$ yield ratio
(Fig. \ref{fig:jpsi_psip_yield}, bottom) is consistent with that observed in
other experiments. 
Fig. \ref{fig:psip_jpsi_sqrts} shows the collision energy dependence
of the $\psip/(\jpsi)$ yield ratio in light fixed target experiments
and \pp or $p+\bar{p}$ colliders. 
In this figure, the ratios from $p+\bar{p}$
experiments were calculated using the reported \jpsi and \psip cross
sections for $p_T>5$ \gevc together with their point-to-point
uncorrelated uncertainties\footnote{This may be an overestimate of the
  systematic errors, given that a good fraction of the \jpsi and \psip
  yields may be correlated.}.  The $B$ meson decay contribution was
removed from the \jpsi and \psip yields, in the case of the CDF
experiment.
Only E705 has broad coverage
($-0.1<x_F<0.5$). The other experiments in this figure have a rapidity
coverage of $|y|<0.6$.  A weak trend of increasing $\psip/(\jpsi)$
yield ratio for higher collision energy
(Fig. \ref{fig:psip_jpsi_sqrts}) and for higher \pt
(Fig. \ref{fig:jpsi_psip_yield}) can be observed.  
As mentioned
earlier, the \psip feed-down fraction of (9.7 $\pm$ 2.4)\% is in
agreement with the world average of $(8.1 \pm 0.3)$\% calculated in
\cite{Faccioli:2008ir}.

\begin{figure}
  \centering
 \includegraphics[width=1.0\linewidth]{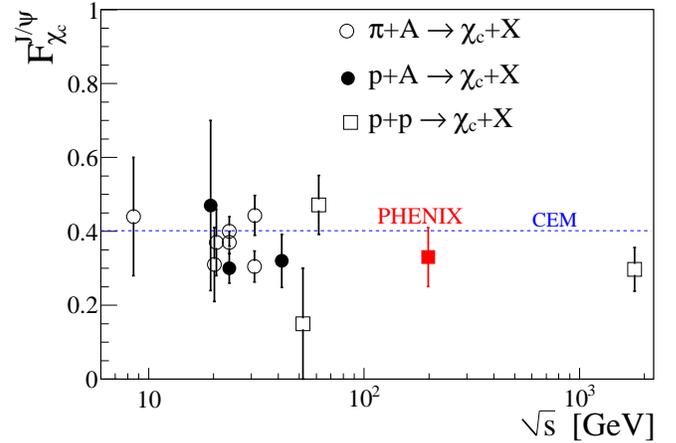}
  \caption{\label{fig:chic_feed_down} (Color online)Collision energy
    dependence of \chic feed-down to \jpsi measured in \pp and $p$+A
    collisions 
    \cite{Hahn:1984tz,Antoniazzi:1992iv,Alexopoulos:1999wp,Abt:2002vq,Binon:1984in,Lemoigne:1982jc,Koreshev:1996wd,Kourkoumelis:1978mj,Abe:1997yz}
    compared to the CEM calculation \cite{Frawley:2008kk}.} 
\end{figure}

The feed-down fraction obtained from our \mbox{$\chic \rightarrow \jpsi +
\gamma$} measurement  is compared with other
experiments over a broad range of collision energy and $x_F$, as well
as over many different colliding species
(Fig. \ref{fig:chic_feed_down}). The 
value measured in this work, $F_{\chi_c}^{J/\psi}=(32 \pm 9) \%$,
is consistent with the world average of ($25 \pm 5$)\% 
after accounting for $A$ dependencies in the fixed target
experiments \cite{Faccioli:2008ir}.\footnote{The world average was obtained in
  after extrapolating the dependence of the
  estimated path length in nuclear matter for the $p$+A fixed
  target experiments.}

Combining the results of feed down from the \psip and the \chic 
we obtain a total \jpsi feed-down fraction measured
in the midrapidity region of $(42 \pm 9)$\%. 


\subsection{Outcomes for heavy ion collisions.}
\label{sec:outcomes}

The
feed-down fractions from the \psip and \chicb have important implications for survival
rates of charmonium states when either ordinary nuclear matter or high
energy density nuclear matter is involved.
Because of their larger size compared to the \jpsi, excited charmonium
states may have a different breakup cross section in nuclear
matter. This effect can modify the feed-down fractions in $p$+A
collisions. On the other hand, if the \cc is not formed as a color
singlet, it can cross the nuclear matter as a colored preresonant
state \cite{Vogt2002539}. If this were true, the breakup cross section
of \jpsi and \psip should be the same and there would be no modification of
the \psip feed-down fraction in $p$+A collisions, whereas a possible
modification can occur for the \chic since it is expected to be formed
mainly as a color singlet.  Given the large statistical uncertainties
in all measured \mbox{\psip / (\jpsi)} ratios shown in
Figs. \ref{fig:jpsi_psip_yield} and \ref{fig:psip_jpsi_sqrts},
differences between $p+p$ and $p$+A are impossible to see, and
therefore no conclusion about possible cold nuclear matter effects can
be made at this time. The same is true for the \chic feed-down
fraction in Fig. \ref{fig:chic_feed_down}. Higher precision
measurements of charmonium states in \pp and $d$+Au collisions in the
future may allow an improved determination of these possible cold
nuclear matter effects on the feed-down fraction.

The behavior of charmonium states in the high density, hot nuclear
matter created in heavy ion collisions, has long been of
interest\cite{Matsui:1986dk}. Spectral function computations
\cite{PhysRevLett.99.211602} indicate that the \chic and \psip states
should dissociate at a lower temperature in hot nuclear matter,
due to color screening, than the \jpsib.  One of the most important
implications of the observed feed-down fractions is that the complete
dissociation of the \chic and \psip states would lead to a (42 $\pm$ 9)\% \jpsi
suppression. The measured nuclear modification
factor of \jpsi mesons in central Au+Au collisions at \full
\cite{Adare:2006ns} implies a suppression of (74 $\pm$ 6)\% at
midrapidity and (84 $\pm$ 6)\% at forward rapidity. Hence, the
complete dissociation of the excited states of charmonium and the
associated loss of the \jpsi yield cannot completely explain its
suppression observed in Au+Au collisions. Cold nuclear matter effects
and the possible dissociation of direct \jpsi by color-screening could
presumably account for the remaining suppression.

\section{Summary and Conclusions}

In conclusion, we have measured the yields of the three most important
charmonium states in \pp collisions at \fullb, where gluon fusion is
expected to be the dominant production process.  The rapidity
dependence of \jpsi supports the use of CTEQ6M to describe the gluon
distribution in protons. The inclusive \jpsi yield is in agreement
with current models which involve a initial formation of colored
charmonium states, as in the CEM or the color octet states of
the NRQCD models. The inclusive \jpsi yield observed at midrapidity is
composed of \mbox{$9.6 \pm 2.4$\%} of \psip decays and \mbox{$32 \pm
  9$\%} of \chic decays.  This result is in agreement with what was
observed in other experiments. Given the current large statistical
uncertainties, no conclusion can be made about collision energy or \pt
dependence of these fractions.  Finally, this \jpsi cross section
measurement and feed-down fractions will play an important role in
current studies of cold nuclear matter and the hot, dense matter
formed in heavy ion collisions.

\section*{Acknowledgments}
\label{sec:ack}


We thank the staff of the Collider-Accelerator and Physics
Departments at Brookhaven National Laboratory and the staff of
the other PHENIX participating institutions for their vital
contributions.  
We also thank Jean Philippe Lansberg, Mathias Butensch\"on and 
Ramona Vogt for the valuable CSM, NRQCD, CEM and FONLL 
calculations in the PHENIX acceptance.
We acknowledge support from the Office of Nuclear Physics in the
Office of Science of the Department of Energy,
the National Science Foundation, 
a sponsored research grant from Renaissance Technologies LLC, 
Abilene Christian University Research Council, 
Research Foundation of SUNY, 
and Dean of the College of Arts and Sciences, Vanderbilt University 
(U.S.A),
Ministry of Education, Culture, Sports, Science, and Technology
and the Japan Society for the Promotion of Science (Japan),
Conselho Nacional de Desenvolvimento Cient\'{\i}fico e
Tecnol{\'o}gico and Funda\c c{\~a}o de Amparo {\`a} Pesquisa do
Estado de S{\~a}o Paulo (Brazil),
Natural Science Foundation of China (P.~R.~China),
Ministry of Education, Youth and Sports (Czech Republic),
Centre National de la Recherche Scientifique, Commissariat
{\`a} l'{\'E}nergie Atomique, and Institut National de Physique
Nucl{\'e}aire et de Physique des Particules (France),
Ministry of Industry, Science and Tekhnologies,
Bundesministerium f\"ur Bildung und Forschung, Deutscher
Akademischer Austausch Dienst, and Alexander von Humboldt Stiftung (Germany),
Hungarian National Science Fund, OTKA (Hungary), 
Department of Atomic Energy and Department of Science and Technology (India),
Israel Science Foundation (Israel), 
National Research Foundation and WCU program of the 
Ministry Education Science and Technology (Korea),
Ministry of Education and Science, Russian Academy of Sciences,
Federal Agency of Atomic Energy (Russia),
VR and the Wallenberg Foundation (Sweden), 
the U.S. Civilian Research and Development Foundation for the
Independent States of the Former Soviet Union, 
the US-Hungarian Fulbright Foundation for Educational Exchange,
and the US-Israel Binational Science Foundation.

\appendix*

\section{Data Tables}

\begingroup \squeezetable

\begin{table}[!ht]
  \caption{\label{tab:jpsi_yield}\jpsi differential cross section
    in the midrapidity region ($|y|<0.35$) followed
    by point-to-point uncorrelated (statistical and uncorrelated
    systematic uncertainties) and correlated systematic
    uncertainties. Global uncertainty is 10\%.}
  \begin{ruledtabular}\begin{tabular}{lcccr}
\pt [\gevc] & \multicolumn{3}{c}{$\frac{1}{2\pi p_T} 
\frac{B_{ee}d^2\sigma_{J/\psi}}{dydp_T}
    [nb/(\gevc)^2]$}\\\hline
 & value & uncor. & corr. \\
    \hline
    0-0.25 & 4.9 & 0.5 &0.6 &  \\
    0.25-0.5 & 3.9 & 0.3 &0.5 & \\
    0.5-0.75 & 3.9 & 0.2 &0.5 & \\
    0.75-1 & 3.5 & 0.2 &0.4 & \\
    1-1.25 & 3.19 & 0.17 &0.38 & \\
    1.25-1.5 & 2.21 & 0.14 &0.27 & \\
    1.5-1.75 & 1.69 & 0.12 &0.2 & \\
    1.75-2 & 1.42 & 0.1 & 0.17 & \\
    2-2.25 & 95 & 8 & 12 & $\times10^{-2}$\\
    2.25-2.5 & 66 & 7 & 8 & $\times10^{-2}$\\
    2.5-2.75 & 56 & 6 & 7 & $\times10^{-2}$\\
    2.75-3 & 37 & 5 & 5 & $\times10^{-2}$\\
    3-3.25 & 29 & 4 & 4 & $\times10^{-2}$\\
    3.25-3.5 & 19.9 & $^{+3.5}_{-3.6}$  & 2.4 & $\times10^{-2}$\\
    3.5-3.75 & 13.6 & 2.8 & 1.7 & $\times10^{-2}$\\
    3.75-4 & 10.6 & 2.5  & 1.3 & $\times10^{-2}$\\
    4-4.25 & 6.89 & $^{+2.3}_{-2.2}$  & 0.8 & $\times10^{-2}$\\
    4.25-4.5 & 4.4 & $^{+1.7}_{-1.6}$  & 0.5 & $\times10^{-2}$\\
    4.5-4.75 & 4.7 & $^{+1.5}_{-1.4}$  & 0.6 & $\times10^{-2}$\\
    4.75-5 & 3.1 & 1.2  & 0.4 & $\times10^{-2}$\\
    5-6 & 1.35 & $^{+0.31}_{-0.32}$  & $^{+0.17}_{-0.18}$ & $\times10^{-2}$\\
    6-7 & 4.1 & 1.3 & $^{+0.5}_{-0.6}$ & $\times10^{-3}$\\
    7-8 & 1.6 & 0.7 & 0.2 & $\times10^{-3}$\\
    8-9 & 0.37 & $^{+0.37}_{-0.22}$ & $^{+0.05}_{-0.06}$ & $\times10^{-3}$\\
  \end{tabular}\end{ruledtabular}
\end{table}

\begin{table}[!ht]
  \caption{\label{tab:psip_yield}\psip differential cross section at
    $|\eta|<$0.35 followed by point-to-point uncorrelated (statistical
    and uncorrelated systematic uncertainties) and correlated
    systematic uncertainties. Global uncertainty is 10\%.}
  \begin{ruledtabular}\begin{tabular}{lccc}
    \pt [\gevc] & \multicolumn{3}{c}{$\frac{1}{2\pi p_T}
    \frac{B_{ee}d^2\sigma_{\psi'}}{dy dp_T}$ [pb/$(\gevc)^2$]}\\\hline
    & value & uncor. & corr. \\\hline
    0-1 & 67 & 20  & 9  \\
    1-2 & 40 & 11  & $^{+7}_{-6}$  \\
    2-3 & 15 & 6  & 3  \\
    3-5 & 2.7 & $^{+2.5}_{-1.5}$  & 0.5  \\
    5-7 & \multicolumn{3}{c}{$<$2.25 (90\% CL)}\\
    0-5 & 95 & 20 & $^{+17}_{-15}$\\
  \end{tabular}\end{ruledtabular}
\end{table}

\begin{table}[!ht]
  \caption{\label{tab:jpsi_rap}Rapidity dependence of \jpsi yield
    followed by point-to-point uncorrelated (statistical and uncorrelated
    systematic uncertainties) and correlated systematic
    uncertainties. Global uncertainty is 10\%.}
  \begin{ruledtabular}\begin{tabular}{lccl}
    rapidity &
    \multicolumn{3}{c}{$B_{ll}\frac{d\sigma_{J/\psi}}{dy}[nb]$}\\\hline
    & value & uncor. & corr. \\\hline
  -2.325 & 10.9 & 1.9 & 1.0 \\
  -2.075 & 17.6 & 0.5 & 1.5 \\
  -1.825 & 24.4 & 0.4 & 1.9 \\
  -1.575 & 31.5 & 0.5 & 2.2 \\
  -1.325 & 41.2 & 1.1 & 5.3 \\
  -0.3   & 49.0 & 2.1 & 5.4 \\
  0.0    & 45.6 & 1.6 & 5.0 \\
  0.3   & 46.1 & 1.9 & 5.1 \\
  1.325 & 40.7 & 1.2 & 5.7 \\
  1.575 & 33.6 & 0.7 & 3.0 \\
  1.825 & 25.6 & 0.4 & 2.5 \\
  2.075 & 18.9 & 0.4 & 1.9 \\
  2.325 & 13.9 & 0.9 & 1.4 \\
  \end{tabular}\end{ruledtabular}
\end{table}

\begin{table}[!ht]
  \caption{\label{tab:psip_jpsi_ratio} \psip / (\jpsi) dielectron yield
    ratio measured at $|\eta|<$0.35 followed by point-to-point
    uncorrelated (statistical and uncorrelated
    systematic uncertainties) and correlated systematic uncertainties.}
  \centering
  \begin{ruledtabular}\begin{tabular}{lccc}
    \pt [\gevc] &
    
\multicolumn{3}{c}{$\frac{B_{ee}^{\psi'}\sigma_{\psi'}}{B_{ee}^{J/\psi}\sigma_
{J/\psi}}$
    [\%]}\\\hline
    & value & uncor. & corr. \\\hline
    0-1 & 1.69& 0.51  & $^{+0.12}_{-0.11}$ \\
    1-2 & 1.96& 0.53  & $^{+0.23}_{-0.15}$ \\
    2-3 & 2.3 & 1.0  & $^{+0.5}_{-0.3}$ \\
    3-5 & 3.4 & $^{+2.0}_{-2.1}$  & $^{+0.5}_{-0.4}$ \\
    5-7 & \multicolumn{3}{c}{$<$38  (90\% CL)}\\
    0-5 & 2.1 & 0.5 & \\
  \end{tabular}\end{ruledtabular}
\end{table}

\begin{table}[!ht]
  \caption{\label{tab:jpsi_yield_forward}Di-muon \jpsi yield in the
    forward rapidity region (1.2$<|y|<$2.4) followed by point-to-point
    uncorrelated (statistical and uncorrelated
    systematic uncertainties) and correlated systematic
    uncertainties. Global uncertainty is 10\%.}
  \begin{ruledtabular}\begin{tabular}{lcccr}
    \pt [\gevc] &
    \multicolumn{3}{c}{$\frac{1}{2\pi p_T} 
\frac{B_{\mu\mu}d^2\sigma_{J/\psi}}{dydp_T}
    [nb/(\gevc)^2]$}\\\hline
  & value & uncor. & corr. \\\hline
  0.125 & 3.49 & 0.14 & 0.50 & \\
  0.375 & 3.28 & 0.08 & 0.49 & \\
  0.625 & 2.85 & 0.06 & 0.40 & \\
  0.875 & 2.43 & 0.05 & 0.16 & \\
  1.125 & 2.04 & 0.04 & 0.18 & \\
  1.375 & 1.57 & 0.03 & 0.16 & \\
  1.625 & 1.194 & 0.024 & 0.12 & \\
  1.875 & 85.1 & 1.9 & 7.9 & $\times 10^{-2}$\\
  2.125 & 63.8 & 1.6 & 5.6 & $\times 10^{-2}$\\
  2.375 & 46.8 & 1.2 & 3.7 & $\times 10^{-2}$\\
  2.625 & 31.5 & 0.9 & 2.5 & $\times 10^{-2}$\\
  2.875 & 22.1 & 0.7 & 1.7 & $\times 10^{-2}$\\
  3.125 & 15.3 & 0.6 & 1.1 & $\times 10^{-2}$\\
  3.375 & 11.1 & 0.5 & 0.8 & $\times 10^{-2}$\\
  3.625 & 7.7 & 0.4 & 0.6  & $\times 10^{-2}$\\
  3.875 & 5.53 & 0.27 & 0.37  & $\times 10^{-2}$\\
  4.125 & 3.28 & 0.21 & 0.23  & $\times 10^{-2}$\\
  4.375 & 2.26 & 0.16 & 0.15  & $\times 10^{-2}$\\
  4.625 & 1.45 & 0.13 & 0.10  & $\times 10^{-2}$\\
  4.875 & 1.06 & 0.11 & 0.08  & $\times 10^{-2}$\\
  5.125 & 1.02 & 0.10 & 0.07  & $\times 10^{-2}$\\
  5.375 & 4.3 & 0.6 & 0.4  & $\times 10^{-3}$\\
  5.625 & 2.9 & 0.4 & 0.2  & $\times 10^{-3}$\\
  5.875 & 2.4 & 0.4 & 0.2  & $\times 10^{-3}$\\
  6.25 & 1.02 & 0.15 & 0.09  & $\times 10^{-3}$\\
  6.75 & 0.73 & 0.13 & 0.06  & $\times 10^{-3}$\\
  7.5 & 0.13 & 0.04 & 0.012  & $\times 10^{-3}$\\
  \end{tabular}\end{ruledtabular}
\end{table}

\endgroup

\clearpage


\end{document}